\documentclass[aps,prd,superscriptaddress,preprintnumbers,floatfix,nofootinbib,notitlepage,showkeys,showpacs, twocolumn]{revtex4-1}

\usepackage[utf8]{inputenc}

\usepackage{graphicx}
\usepackage{hyperref}
\usepackage{latexsym}
\usepackage{amsmath}
\usepackage{amssymb}
\usepackage{bbm}
\usepackage{ulem}
\usepackage{pdfsync}
\usepackage{epsfig}
\usepackage{epstopdf}
\usepackage{subfigure}
\usepackage{xcolor}
\usepackage{comment}
\usepackage{slashed}
\usepackage{multirow}
\usepackage{xfrac}

\newcommand{\avg}[1]{\langle #1 \rangle}

\newcommand{\MZ}[1]{\textcolor{red}{[{\bf MZ}: #1]}}
\newcommand{\htot}{h_t^2}
\begin{document}

\title{The distribution of the gravitational-wave background from supermassive black holes}
\author{Gabriela Sato-Polito}
\email{gsatopolito@ias.edu}
\affiliation{School of Natural Sciences, Institute for Advanced Study, Princeton, NJ 08540, United States}

\author{Matias Zaldarriaga}
\affiliation{School of Natural Sciences, Institute for Advanced Study, Princeton, NJ 08540, United States}

\begin{abstract}
The recent detection of gravitational waves (GWs) by pulsar timing array (PTA) collaborations spurred a variety of questions regarding the origin of the signal and the properties of its sources. The amplitude of a GW background produced by inspiralling supermassive black holes (SMBHs) can be predicted in a relatively robust manner from the present-day merged remnants, observed as single SMBHs at the centers of galaxies, but falls short of the signal measured by PTAs by a significant amount, requiring equal mass mergers, extremely short delay times, and no accretion in order to achieve a modest consistency. In this work, we revisit NANOGrav's 15-yr data set and reassess the aforementioned discrepancy using the full spectral information captured by PTA data. As previously noted in the literature, the discrete number of point sources contributing to the background may lead to deviations in the observed spectrum relative to the average ($h^2_c \propto f^{-4/3}$) due to Poisson fluctuations, providing additional information about the source population beyond the background amplitude. We derive a simple expression for the characteristic strain distribution given a SMBH model, which is generally applicable regardless of the method used to model the black hole population. We then refit the NANOGrav free spectrum using a minimal model based on the local mass function, showing that the current GW measurement requires roughly $\sim 10$ times more black holes than suggested by local observations and disfavors mass functions dominated by few very heavy sources, with the typical mass that contributes to the background $\lesssim 10^{10}M_{\odot}$. Given the range of SMBH models found to be consistent with the isotropic background, we address what is the typical number sources that would be individually detectable, given the current sensitivity.
\end{abstract}

\maketitle
\section{Introduction}
Low-frequency gravitational wave measurements through pulsar timing arrays (PTAs) represent a new window into a largely unexplored part of the Universe. The recent first detection~\cite{NANOGrav_stoc, EPTA:2023fyk, PPTA, CPTA} of a stochastic gravitational-wave background (SGWB) has prompted the question of whether the observed signal is consistent with expectations based on a wealth of simulations and electromagnetic observations of supermassive black holes~\cite{NANOGrav_SMBH, NANOGrav_howto,EPTA_SMBH, Ellis:2023oxs, Ellis:2024wdh, Sato-Polito:2023gym}. Given the current GW measurements, an emerging consensus appears to be that the observed signal is relatively high compared to predictions, with most (e.g. Refs.~\cite{NANOGrav_SMBH} and \cite{EPTA_SMBH}) suggesting a high but consistent value for the observed SGWB when compared to supermassive black hole (SMBH) models.

A variety of methods have been employed to model the SGWB produced by inspiralling SMBHs \cite{rajagopal_ultra_1995, Wyithe:2002ep, Phinney:2001di, Sesana:2008mz}. Most models that are currently compared to PTA observations are based on quantifying the merger rate density of halos or galaxies over cosmic history. This can be achieved via self-consistent predictions of the merger rate from N-body simulations (see, e.g., Refs.~\cite{Ravi:2012bz, Kelley:2016gse, 2022MNRAS.509.3488I}) or based on the extended Press-Schechter formalism~\cite{Sesana:2008mz, Ellis:2023dgf}. Another option is a more empirical route of parametrizing the galaxy mass function and various relevant properties that determine the associated SMBH merger history (such as the time delay, pair fraction, etc.) and fitting them to available observations~\cite{Sesana:2012ak, Chen:2018znx, Simon:2023dyi}.

As noted by Ref.~\cite{Phinney:2001di}, the amplitude of the SGWB produced by merging black holes can be directly related to the mass function of the remnant population today. Using local scaling relations between supermassive black holes combined with galaxy catalogs, Ref.~\cite{Sato-Polito:2023gym} showed that the resulting SGWB falls short of the measured background by a significant amount, requiring multiple merger events at very low redshifts, with equal mass companions, and with negligible accretion to achieve the central value reported by the NANOGrav collaboration (with equivalent results found by EPTA+InPTA and PPTA). 

While Ref.~\cite{Sato-Polito:2023gym} focused on the power law fit assuming a fixed index consistent with the mean spectrum predicted for a SMBH population \cite{Phinney:2001di}, here we explore the full spectral information. We describe the total characteristic strain in each frequency bin captured by a given PTA as the sum of the GW emission from each discrete source ($h^2_s$), which can lead to deviations in the spectrum relative to the mean due to Poisson fluctuations in a given realization of the observed GW sky \cite{Sesana:2008mz, NANOGrav_discrete}. That is, given a set of model parameters $\vec{\theta}$ that describes the merger of SMBHBs, there is a distribution $p(h^2_c(f)|\vec{\theta})$ of the observed GWB and we present an analytical prediction of this distribution.

To do so, we follow the literature on the confusion limit (unresolved background) of point sources first developed in the context of radio observations \cite{Scheuer_1957,1974ApJ...188..279C} and subsequently adapted into studies across the electromagnetic spectrum (see, e.g., \cite{Lee:2008fm, 2010MNRAS.409..109G, Breysse:2016szq}). We recast the description of the characteristic strain distribution in terms of a ``luminosity function" for GW emitters $dN/dh^2_s$ (perhaps more accurately called a flux, but we use the term luminosity function throughout). We then present an analytical description of the characteristic strain distribution as a function of frequency, given a SMBH model, which only depends on its luminosity function. The resulting distribution is identical to the standard Monte Carlo approach adopted throughout the literature (e.g. \cite{Sesana:2008mz, Ravi:2012bz}), but can be computed directly. We emphasize that, while we employ one particular approach for estimating the SMBH merger rate and therefore the GW luminosity function, the method we present for computing $p(h^2_c(f)|\vec{\theta})$ is general. For instance, if a simulation-based approach is preferred, one need only measure $dN/dh^2_s$ from the simulations and $p(h^2_c(f)|\vec{\theta})$ can be computed in the same manner.


We then use the model outlined above to fit NANOGrav's 15 yr free-spectrum posterior \cite{NANOGrav_stoc, Lamb_freespec}. Our results are consistent with earlier work presented in Ref.~\cite{Sato-Polito:2023gym}, finding that a higher black hole abundance or higher masses are required to fit the NANOGrav data, in tension with estimates of the present-day SMBH mass function derived from local scaling relations and galaxy catalogs. However, the addition of spectral information disfavors models in which a small number of bright sources  dominate (i.e. in which the typical mass that contributes to the background is large). This conclusion is primarily driven by the fact that for backgrounds dominated by an extremely small number of very massive sources, deviations in the spectrum away from the average power law ($h^2_c \propto f^{-4/3}$) begin to impact the lowest frequencies captured by the data set.

We also emphasize that the results presented in this work are somewhat consistent with those from the NANOGrav and EPTA+InPTA collaborations \cite{NANOGrav_SMBH, EPTA_SMBH}, which also find a relatively high amplitude for the GWB, captured through a preference for a high amplitude of the galaxy stellar mass function or normalization of the merger rate density, a time delay posterior peaked at 0, and a slight preference for models in which the typical number of sources is large (e.g. mild preference for low scatter in the scaling relation). The analysis shown in this work accounts for the full distribution $p(h^2_c(f)|\vec{\theta})$, while Ref.~\cite{NANOGrav_SMBH} assumes a Gaussian centered on the median of $h^2_c$, while Ref.~\cite{EPTA_SMBH} assumes a deterministic model. However, this is unlikely to lead to a significant difference in conclusions, since the distribution only becomes appreciably non-Gaussian in the regime of low number of sources. Conversely, we only include a relatively simple astrophysical model, ignoring binary interactions with its environment and accretion. However, the former will only significantly affect the lowest frequencies and, since we follow the approach from Ref.~\cite{Phinney:2001di} which anchors the SGWB prediction on the present-day black hole remnant population, the inclusion of accretion can only reduce the predicted background, exacerbating the aforementioned discrepancy.

Sufficiently loud sources are expected to be individually detectable \cite{Sesana:2008xk, Rosado:2015epa, Kelley:2017vox, Becsy:2022pnr}. We therefore investigate what a high amplitude GWB implies for the detectability of individual sources given the current sensitivity reported by Ref.~\cite{NANOGrav_individual}. Given the region of parameter space that satisfies the isotropic background measurement, we ask: what is the probability that a point source should have been detected? While the answer depends on the unknown redshift distribution of sources, we show that an upper limit can always be derived. We show that the maximum average number of sources (across all frequencies) consistent at a $3\sigma$ CL with the GWB measurement is $\bar{N} \sim 0.1$, while for a SMBH mass function consistent with local estimates and a redshift distribution consistent with simulations, we find that $\bar{N} \sim 10^{-3}$. Hence, while the non-detection of an individual source does not currently add any information regarding the SMBH distribution, we conclude that improving the measurement of the frequency dependence of the GWB may determine whether a single source detection is likely in the near future or not.

This paper is organized as follows. In Sec.~\ref{sec:mean_h2c} we review the standard calculation of the mean characteristic strain spectrum from the local black hole remnant population. In Sec.~\ref{sec:hc_pdf}, we present the derivation of the distribution of characteristic strains and a discussion regarding the impact of the typical number of sources that contribute to the background, as well as the changes induced by multiple merger events. Sec.~\ref{sec:PTA_spectrum} shows the 15-yr NANOGrav spectrum to be fit and the main results are presented in Sec.~\ref{sec:results}, with a discussion regarding both the isotropic background fit and its implications for individual source detection. We conclude in Sec.~\ref{sec:conclusion}. 

\section{Average characteristic strain}\label{sec:mean_h2c}
\begin{figure}
    \centering
    \includegraphics[width=0.42\textwidth]{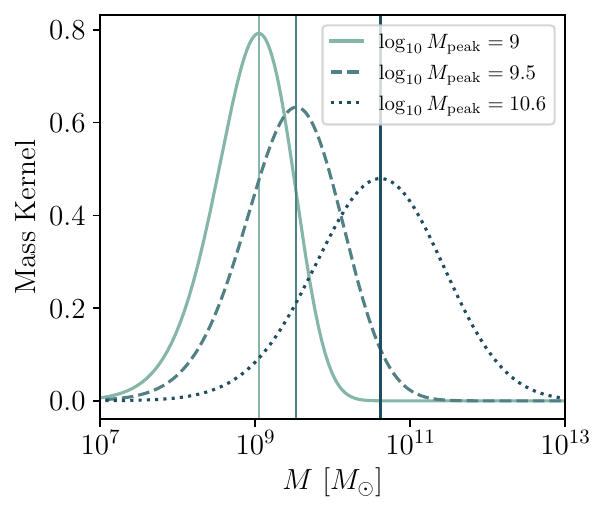}
    \caption{Contribution to the characteristic strain squared per logarithmic black hole mass. Each kernel corresponds to a different model for the mass function and the peak mass is defined as the peak of the kernel, shown in the vertical lines.}
    \label{fig:mass_kernel}
\end{figure}

While standard in the literature, we will restate the usual expression for the stochastic gravitational-wave background produced by a collection of inspiralling SMBHBs \cite{rajagopal_ultra_1995, Phinney:2001di}, which can be written as the sum of the emission from all binaries. As argued in Ref.~\cite{Phinney:2001di}, the black hole merger rate density can be directly related to the present-day ($z=0$) remnant single black hole mass function $\phi$. The average characteristic strain spectrum in the Universe is therefore given by
\begin{equation}
    h^2_c(f) =\frac{4 G}{\pi f^2} \avg{(1+z)^{-1/3}}\int dM\ M \phi(M) \avg{\epsilon}(M),
    \label{eq:h2c_mean}
\end{equation}
where $\phi$ is a function of total mass $M$, and we have defined the redshift and mass ratio averages
\begin{equation}
    \avg{(1+z)^{-1/3}} = \int dz \frac{p_z(z)}{(1+z)^{1/3}}
\end{equation}
and
\begin{equation}
    \avg{\epsilon}(M)= \int dq\ p_q(q) \epsilon(M, q),
\end{equation}
where $\epsilon \equiv \frac{\eta}{3} \frac{(G M \pi f)^{2/3}}{c^2}$ is a GW efficiency parameter, $\eta=q/(1+q)^2$ is the symmetric mass ratio, and $f$ the observed GW frequency. The mean characteristic strain $h^2_c$ receives most of its contribution from a characteristic mass $M_{\rm peak}$, as shown by the mass kernel in Fig.~\ref{fig:mass_kernel}. 

As in Ref.~\cite{Sato-Polito:2023gym}, we choose the following parametrizations for the redshift and mass-ratio distribution functions
\begin{equation}
    p_z(z) = z^\gamma e^{-z/z_*}, \quad \text{and} \quad p_q(q) = q^{\delta},
    \label{eq:pz_pq}
\end{equation}
with fiducial values of $\gamma=0.5$, $z_*=0.3$, and $\delta=-1$, an assumed minimum value of $q_{\rm min}=0.1$, and where both functions are then normalized to unity. These fiducial values are chosen in order to roughly reproduce the simulation-based results presented in Ref.~\cite{Kelley:2016gse} (see discussion in Ref.~\cite{Sato-Polito:2023gym}).

The single supermassive black hole mass function $\phi$ at $z=0$ can be estimated from local scaling relations that relate the black hole mass to the properties $X$ of the host galaxy, i.e.
\begin{equation}
    \log_{10}M = a_{\bullet} + b_{\bullet}\log_{10} X.
    \label{eq:M-X}
\end{equation}
This scaling relation can then be combined with galaxy catalogs which measure the number density of galaxies as a function of the property $X$, where we adopt the velocity dispersion as our fiducial proxy for SMBH mass. The velocity dispersion function (VDF) is parameterized as
\begin{equation}
    \phi(\sigma) d\sigma = \phi_{*} \left(\frac{\sigma}{\sigma_*}\right)^{\alpha} \frac{e^{-(\sigma/\sigma_*)^{\beta}}}{\Gamma(\alpha/\beta)} \beta \frac{d\sigma}{\sigma}.
    \label{eq:sigma_function}
\end{equation}
In the presence of scatter between black hole mass and galaxy property, we compute the SMBH mass function as
\begin{equation}
    \phi(M) = \int d\sigma \frac{p(\log_{10}M|\log_{10}\sigma)}{M \log(10)} \phi(\sigma),
    \label{eq:MF_scatter}
\end{equation}
where we assume $p$ to be log-normal
\begin{equation}
\begin{split}
    p(\log_{10}&M|\log_{10}\sigma) = \frac{1}{\sqrt{2\pi} \epsilon_0} \\ &\times \exp\left[ -\frac{1}{2}\left(\frac{\log_{10}M - a_{\bullet}-b_{\bullet}\log_{10}\sigma}{\epsilon_0}\right)^2 \right].
\end{split}
\end{equation}

Similarly to Ref.~\cite{Sato-Polito:2023gym}, we choose a fiducial velocity dispersion function consistent with the measurement from Ref.~\cite{2010MNRAS.404.2087B} from SDSS, fit to all galaxies with $\sigma>125$km s$^{-1}$, which leads to measured parameter values of $\phi_*=(2.61 \pm 0.16)\times 10^{-2}$Mpc$^{-3}$, $\sigma_*=159.6\pm1.5$ km s$^{-1}$, $\alpha=0.41\pm 0.02$, and $\beta=2.59\pm 0.04$. For the scaling relation, we use $X=\sigma/200$km s$^{-1}$ in Eq.~\ref{eq:M-X} and the $M-\sigma$ relation from Ref.~\cite{mcconnell_ma}, which corresponds to $a_{\bullet}= 8.32\pm 0.05$, $b_{\bullet}= 5.64\pm 0.32$, and $\epsilon_0=0.38$. As demonstrated in Ref.~\cite{Sato-Polito:2023gym}, the particular choice of BH mass proxy (whether the velocity dispersion or the stellar mass) for the scaling relation, or the galaxy survey used does not change the gravitational wave prediction in an appreciable way.

\section{Characteristic strain distribution}\label{sec:hc}
\subsection{Distribution Function}\label{sec:hc_pdf}
The result presented in the previous section is an accurate description of the mean characteristic strain spectrum in the Universe as seen by an observer today. The observed GWB, however, is produced by a discrete population of SMBHBs, which can be understood as a single realization of the underlying distribution of characteristic strains and may differ substantially from the mean. The total characteristic strain is given by the sum over all sources, labelled $s$, such that in each frequency bin we have that
\begin{equation}
    \htot (f)=\sum_s h_s^2 (f).
\end{equation}
We will assume that sources are distributed according to a luminosity function $\frac{dN}{dh_s^2}$, and the dependence on the frequency is suppressed for simplicity, but we stress that the quantities discussed correspond to values within a discrete frequency bin. More explicitly, the luminosity function is $\frac{dN}{d h^2_s}(f) = \int_{\Delta f} \frac{dN}{d h^2_s d\log f} \approx \frac{\Delta f}{f} \frac{dN}{dh^2_s d\log f}$. In terms of this luminosity function, the average number of sources ${\bar N}$ and average power $\bar{h_t^2}$ are given by:
\begin{equation}
\begin{split}
    \bar{N}=& \int dh_s^2 \frac{dN}{dh_s^2} \\
    \bar{\htot} =& \int dh_s^2 \frac{dN}{dh_s^2} h_s^2.
\end{split}
\end{equation}
The mean characteristic strain in Eq.~\ref{eq:h2c_mean} is therefore $h^2_c= \bar{\htot}$. 

Following the literature on the confusion limit, the probability distribution of characteristic strains $h^2_t$ can be written as the probability $P_{\rm Poiss}(N_s|\bar{N})$ of there being $N_s$ sources and the probability $P(h^2_t|N_s)$ that those $N_s$ sources produce an aggregate spectrum $h^2_t$, summed over all values of $N_s$. That is,
\begin{equation}
    P(h^2_t) = \sum_{N_s=0}^{\infty} P_{\rm Poiss}(N_s|\bar{N})P(h^2_t|N_s).
    \label{eq:Ph2c_tdomain}
\end{equation}
For the case of $N_s=0$, the probability is of course given by $P(h^2_t|N_s=0) = \delta_D(h^2_t)$. If there is only one source, we can compute the probability in terms of the luminosity function
\begin{equation}
P(\htot|N_s=1) = \frac{1}{\bar N} \frac{dN}{dh_s^2},
\end{equation}
and, since $\htot$ is a sum of the power of each source which are assumed to be independent, $P(\htot|N_s)$ is equal to $N_s$ convolutions of $P(\htot|N_s=1)$ with itself. This therefore suggests that Eq.~\ref{eq:Ph2c_tdomain} can be more conveniently computed in the Fourier domain. Taking the Fourier transform of the distribution of $\htot$, we find
\begin{equation}  
\hat P(\omega)=\sum_{N_s=0}^{\infty} P(N_s|\bar N) \hat{P}(\omega | N_s),
\label{eq:Ph2c_fdomain}
\end{equation}
where $\hat{P}(\omega | N_s)$ is the Fourier transform of $P(\htot|N_s)$. The $N_s$ convolutions of $P(\htot|N_s=1)$ can now be written as the Fourier transform of this distribution to the $N_s$ power, i.e.
\begin{equation}
    \hat{P}(\omega | N_s) = [\hat{P}(\omega | N_s=1)]^{N_s}.
\end{equation}
Substituting this result into Eq.~\ref{eq:Ph2c_fdomain} leads to the more convenient expression
\begin{equation}
\begin{split}
    \hat P(\omega)=&e^{-\bar{N}} \sum_{N_s=0}^{\infty} \frac{[\bar{N} \hat{P}(\omega | N_s=1)]^{N_s}}{N_s!}\\
    =& \exp\left\{\bar{N}\hat{P}(\omega | N_s=1) - \bar{N}\right\} \equiv \exp[u(\omega)]
\end{split}
\end{equation}
where
\begin{align}
    {\rm Re}[u(\omega)] =& \int dh_s^2 \frac{dN}{dh_s^2} (\cos(\omega h_s^2)-1) \\
    {\rm Im}[u(\omega)] =& \int dh_s^2 \frac{dN}{dh_s^2} \sin(\omega h_s^2).
\end{align}
The probability distribution for $h_t^2$ can then be recovered by taking the inverse Fourier transform of the expression above
\begin{equation}
P(h_t^2)= \int\ d\omega\  e^{i \omega \htot} \hat P(\omega).
\end{equation}
The results presented in this work are obtained by numerically computing the expressions above, but we provide an in-depth discussion in App.~\ref{app:dist_power} of analytical results that we derive assuming power-law luminosity functions. As we will show in the subsequent discussion in this Section, the power-law approximation is a relatively good one, but we nevertheless opt to include the full shape of the luminosity function in the main results.

The only remaining quantity that must be computed is the distribution of characteristic strains for a single source. This can be directly derived from the comoving number density of mergers, assuming that all SMBHs in the PTA band are in circular orbits, evolving primarily due to GW emission. Similarly to Ref.~\cite{Sesana:2008mz}, we first compute the total number of mergers per logarithmic frequency and binary parameters
\begin{equation}
    \frac{dN}{dz dM dq d\log f} = \frac{dn}{dz dM dq} \frac{dt_r}{d\log f_r} \frac{dz}{dt_r} \frac{dV_c}{dz},
    \label{eq:dNdp}
\end{equation}
where $t_r$ is the coordinate time in the source rest frame, $V_c$ is the comoving volume, and $f_r=(1+z)f$. The frequency evolution is given by
\begin{equation}
    \frac{d\log f_r}{dt_r} = \frac{96 \pi^{8/3}}{5}\left(\frac{G M}{c^3}\right)^{5/3} \eta f_r^{8/3},
    \label{eq:dlogfdt}
\end{equation}
where
\begin{equation}
    \frac{dz}{dt_r} \frac{dV_c}{dz} = (1+z)H(z) \frac{4\pi c \chi^2(z)}{H(z)},
    \label{eq:dVdt}
\end{equation}
where $\chi$ is the comoving radial distance and $H(z)$ the Hubble parameter. The contribution of each binary to the characteristic strain, averaged over inclination and polarization angles, is then given by \cite{2004ApJ...611..623S}
\begin{equation}
    \tilde h_s^2 (M,q,z,f)= \frac{32}{5} \left(\frac{G M}{c^2}\right)^{10/3} \eta^2 \frac{(1+z)^{4/3}}{\chi^2(z)} \left(\frac{\pi f}{c}\right)^{4/3} \frac{f}{\Delta f},
    \label{eq:tildehs2}
\end{equation}
where $\Delta f = 1/T_{\rm obs}$ and $T_{\rm obs}$ is the total observing time.

\begin{figure}[t]
    \centering
    \includegraphics[width=0.45\textwidth]{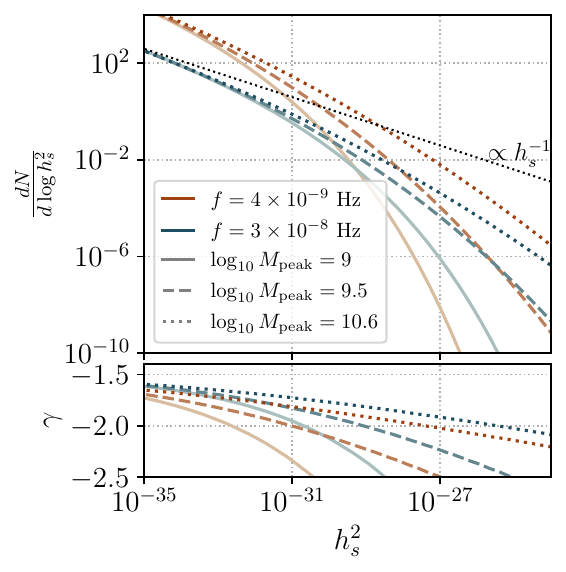}
    \caption{Number of sources per logarithmic characteristic strain for individual sources. The top panel shows the distribution for frequencies corresponding roughly to frequencies of $(10 {\rm yr})^{-1}$ and (yr)$^{-1}$, in orange and blue lines, respectively. The different line styles correspond to different values of peak mass that contributes to the average characteristic strain, obtained in practice by varying the scatter in the $M-\sigma$ relation. The bottom panel shows the power-law index of $dN/dh^2_s$ as a function of $h^2_s$ for each frequency and mass function.}
    \label{fig:LF_gamma}
\end{figure}

The number of sources per emitted GW amplitude $dN/\log h^2_s d\log f$ can be directly computed by changing variables and integrating the number of sources over redshift and mass ratio. That is,
\begin{equation}
\begin{split}
    \frac{dN}{d\log h^2_s d\log f} =& \int d\log M \int dz \int dq\  \delta^D(\log h^2_s - \log \tilde h^2_s) \\ &\times \frac{dN}{d\log M dq dz d\log f} \\
    =& \int d\log M \int dz \int dq\  \delta^D(\log h^2_s - \log\tilde h^2_s) \\ &\times \frac{dn}{d\log M} p_z(z) p_q(q) \frac{dt_r}{d\log f_r} \frac{dz}{dt_r} \frac{dV_c}{dz},
\label{eq:dNdhsdlogf}
\end{split}
\end{equation}
where $\tilde h^2_s \equiv \tilde h^2_s(M,q,z,f)$ is the function given in equation \ref{eq:tildehs2}. The required luminosity function is simply $\frac{dN}{d\log h^2_s}(f) \approx \frac{\Delta f}{f} \frac{dN}{d\log h^2_s d\log f}$.  We can choose to do one of the integrals using the Dirac delta function. It is convenient to do the integral over mass:
\begin{equation}
\begin{split}
\frac{dN}{d\log h^2_s} =&  \int dz \int dq\ \left|\frac{\partial\log M}{\partial \log h^2_s}\right| \frac{dn}{d\log M} p_z(z) p_q(q)  \\ &\times\frac{dt_r}{d\log f_r} \frac{dz}{dt_r} \frac{dV_c}{dz} \frac{\Delta f}{f}
\end{split}
\label{eq:luminosity-function}
\end{equation}
where $\left|{\partial\log  M}/{\partial \log h^2_s}\right|=3/10$ and one has to evaluate the mass function at the mass obtained by solving $\tilde h^2_s(M,q,z,f)=h^2_s$. The dependence on $h_s^2$ in the luminosity function comes only from the dependence of $M$ on $h_s^2$. 

We can understand the basic features of the the luminosity function by examining \ref{eq:luminosity-function}. We first note that the dependence on the mass and mass ratio come through $\tilde h^2_s(M,q,z,f)$ and ${dt_r}/{d\log f_r}$ which in both cases combine through the chirp mass $M_c=M\eta^{3/5}$. Thus we can do the integral over the mass ratio to obtain:
\begin{align}
\frac{dn}{d\log M_c}  =& \int \log M \int dq\ \delta^D(\log M_c -\log(M\eta^{3/5})) \nonumber \\
&\times p_q(q) \frac{dn}{d\log M}  \\
    \frac{dN}{d\log h^2_s} =&  \int dz \  \left|\frac{\partial\log  M_c}{\partial \log h^2_s}\right| \frac{dn}{d\log M_c} p_z(z) \nonumber \\ &\times\frac{dt_r}{d\log f_r} \frac{dz}{dt_r} \frac{dV_c}{dz} \frac{\Delta f}{f}.
\label{eq:luminosity-function-Mc}
\end{align}
For the purpose of understanding the luminosity function, we can keep in mind that the shape of  ${dn}/{d\log M_c}$ is basically the same as that of ${dn}/{d\log M}$.  The volume factor in the integral is:
\[
\frac{dV_c}{d\log f} = \frac{dt_r}{d\log f_r} \frac{dz}{dt_r} \frac{dV_c}{dz} \propto M_c^{-5/3} f^{-8/3}.
\]
As a function of redshift ${dV_c}/{d\log f}$ peaks around redshift two but with our choice of $p_z$,  $p_z{dV_c}/{d\log f}$ peaks at redshift about one.  This is the redshift that determines the mapping between $M$ and $h_s^2$.  A change in the redshift shifts the relationship between the mass and $h_s^2$:
\[
\log h_s^2 = \frac{10}{3}\log M_c + \log \frac{(1+z)^{4/3}}{\chi^2(z)} + {\rm constant}. 
\]
Thus the luminosity function is a superposition of shifted copies of the mass function, shifted by the redshift factor $\log({(1+z)^{4/3}}/{\chi^2(z)})$ weighted $p_z{dV_c}/{d\log f}$. Both functions of $z$ peak at redshifts one to two and thus there is a fairly tight relation between the strain and the mass that contributes the most to that strain. The shape of the luminosity function is roughly given by 
\[
\frac{dN}{d\log h^2_s} \propto M_c^{-5/3} \frac{dn}{d\log M_c} \ \ ; \ \ M_c\propto h_s^{3/5}.
\]
At low values of $h_s^2$, corresponding to low masses, the mass function is approximately constant so ${dN}/{d\log h^2_s}\propto h_s^{-1}$. 

\begin{figure*}[t]
    \centering
    \includegraphics[width=0.85\textwidth]{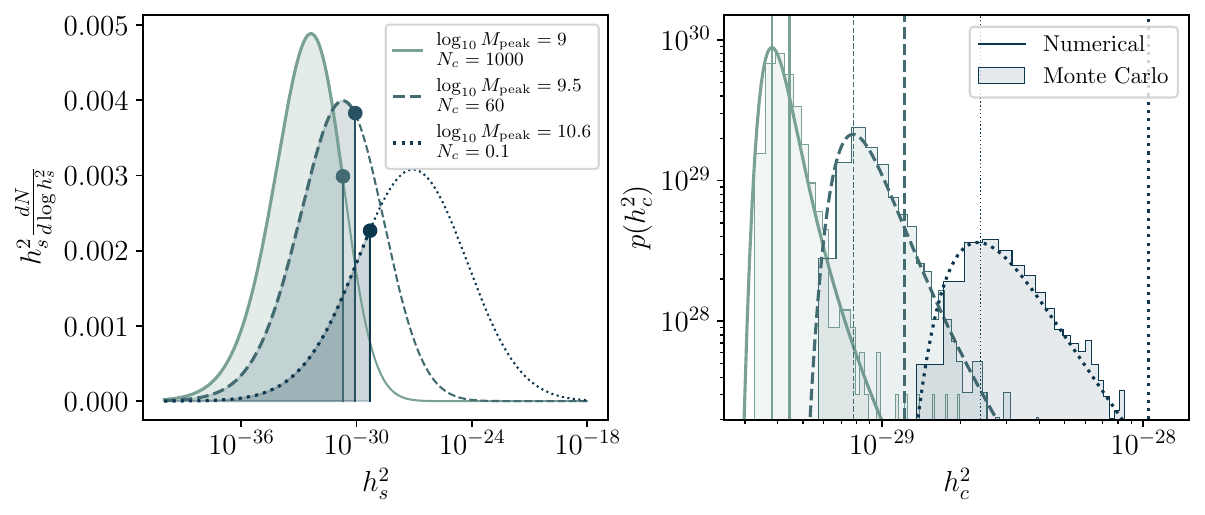}
    \caption{Kernel of the mean characteristic strain for different SMBH models on the left panel and their distributions on the right, all shown for a frequency of $f=4\times 10^{-9}$Hz. The dot on the left panel shows the point $h^2_{s, 1}$, defined in Eq.~\ref{eq:h2one}. The peak of the distribution on the right panel can be estimated via Eq.~\ref{eq:h2peak} (shown in the thin vertical lines), and the integral of the shaded region on the left panel corresponds exactly to the value of the peak. The thick vertical lines show the mean, computed via the full integral of the kernel, showing that the smaller the number of sources that contribute to the background, the more the mean is dominated by the bright (rare) tail of the distribution and differs from the peak of the distribution. We also compare the numerical results presented in this work with the standard Monte Carlo approach, finding excellent agreement.}
    \label{fig:h2c_kernel_pdf}
\end{figure*}

The presence of a characteristic mass that most contributes to the background $M_{\rm peak}$ will lead to a corresponding scale $h_s^2|_{\rm peak}$ on the luminosity function. 
If we think of the luminosity function as a function of  $M_{\rm peak}$, the comoving density of the sources at that peak $\phi_\star$ and the frequency one concludes that the normalization of the luminosity is a very strong function of both frequency and $M_{\rm peak}$,
\[
\frac{dN}{d\log h^2_s} \propto \phi_\star M_{\rm peak}^{-5/3} f^{-11/3}.
\]
The location of this turn over in the luminosity function also scales with $M_{\rm peak}$ and the frequency as a result of the dependence in \ref{eq:tildehs2}
\[
h_s^2|_{\rm peak} \propto M_{\rm peak}^{10/3} f^{7/3}.
\]
Similarly to the mass kernel shown in Fig.~\ref{fig:mass_kernel}, we show the characteristic strain kernel in the left panel of Fig.~\ref{fig:h2c_kernel_pdf}.

Fig.~\ref{fig:LF_gamma} shows the luminosity function of strain amplitudes at different frequencies and for black hole mass functions with different values of peak mass. The shape of this function is clearly related to the shape of the black hole mass function. Changes in the frequency and the peak mass of the black hole mass function shift the luminosity function both vertically and horizontally, as discussed above. The bottom panel shows, in the notation of App.~\ref{app:dist_power}, the power-law index $\gamma$ of the strain distribution $dN/dh^2_s$. We also note that the strain amplitude at which $\gamma=-2$ corresponds exactly to the point $h^2_{s, {\rm peak}}$, since it the power-law index at which the integral of $\htot$ transitions from being dominated by the upper limit to the lower limit of the integral over strain.

The panel on the right of Fig.~\ref{fig:h2c_kernel_pdf} shows the final distribution computed through the approach described in this section and we compare it with the standard Monte Carlo approach of sampling the mass, redshift, and mass ratio distributions. In this case, the mean number of sources within a bin is computed through Eq.~\ref{eq:dNdp} as $\bar{N}_{ijkl} = \frac{d^4N}{d\log f dM dz dq} \Delta \log f_i \Delta M_j \Delta z_k \Delta q_l$. Each of the $N$ sources within a given bin then produces a strain amplitude $h^2_{ijkl} \equiv h^2_s (f_i, M_j, z_k, q_l)$ (from Eq.~\ref{eq:tildehs2}), where $N$ is sampled from a Poisson distribution with mean $\bar{N}_{ijkl}$. We therefore model the characteristic strain through
\begin{equation}
    h^2_c(f_i) = \sum_{jkl} P_{\rm poiss}(N,\bar{N}_{ijkl}) h^2_{ijkl}. 
\end{equation}
Each of the distributions shown in Fig.~\ref{fig:h2c_kernel_pdf} correspond to 1000 realizations of the equation above, demonstrating excellent agreement between the standard Monte Carlo approach and the analytical formulas presented in this work.

\subsection{Characteristic number of sources}
Let us define the characteristic number of sources $N_c$ that contribute to the background at a given frequency bin as
\begin{equation}
N_c(f) \equiv \frac{h_c^2(f)}{h_s^2|_{\rm peak}(f)},
\label{eq:Nc_def}
\end{equation}
where ${h_s^2|_{\rm peak}(f)}$ is the strain of the sources at the peak of the mass kernel of $h_c^2$. Both the total characteristic strain and the strain of a source have a simple scaling with the parameters of the model. The background strain scales as
\begin{equation}
h_c^2(f) \propto \phi_\star M_{\rm peak}^{5/3} f^{-4/3},
\label{eq:h2c_scaling}
\end{equation}
and the source strain
\begin{equation}
h_s^2(f) \propto M_{\rm peak}^{10/3} f^{7/3}.
\end{equation}
The expressions above are slightly modified when changing the scatter in the scaling relation, which adds an additional change in the width of the mass kernel. The characteristic strain then scales with an additional factor of $e^{25/18 \epsilon_0^2 \log^2(10)}$. The expressions above imply that the characteristic number of sources is a very strong function of both frequency and peak mass
\begin{equation}
\begin{split}
{N_c}\approx 0.2 &\times  \left(\frac{\phi_\star}{2.6\ 10^{-2}\ {\rm Mpc}^{-3}} \right) \left(\frac{M_{\rm peak}}{10^{10}M_\odot} \right)^{-5/3} \\ &\times \left(\frac{f}{1/3\ {\rm yr}^{-1}} \right)^{-11/3}.
\end{split}
\end{equation}

Notice that for many values of the parameters, the cross over between $N_c \sim 1$ will happen in the middle of the PTA band, which implies that the smallness of the number of sources will produce a significant effect in the predicted characteristic strain spectrum. Another relevant scenario is to keep $h^2_c \propto \phi_\star M_{\rm peak}^{5/3}$ fixed, since it is the main quantity directly constrained by the PTA data. In that case:
\begin{equation}
\begin{split}
{N_c} \approx 1 \times&  \left(\frac{h_c(f= {\rm yr}^{-1})}{2.4\times 10^{-15}}\right)^2  \left(\frac{M_{\rm peak}}{10^{10}M_\odot}\right)^{-10/3} \\ \times& \left(\frac{f}{1/3\  {\rm yr}^{-1}}\right)^{-11/3}.
\label{eq:Nc_h2c_scaling}
\end{split}
\end{equation}
This is a very strong function of $M_{\rm peak}$ which will allow us to constrain its value, contingent on the shape of the frequency dependence of the spectrum. 

We show in Fig.~\ref{fig:h2c_kernel_pdf} the characteristic strain kernel for different SMBH mass functions, computed by varying the scatter in the scaling relation. The dot in the left panel marks the point after which all contributions to its right are expected to be produced by a single source. That is, the value of $h^2_{s, 1}$ of that point is defined by
\begin{equation}
    \int_{h^2_{s, 1}}^{\infty} dh^2_s \frac{dN}{dh^2_s} = 1,
    \label{eq:h2one}
\end{equation}
and therefore do not typically contribute to the mean characteristic strain $h^2_c$. The mean characteristic strain up to the point $h^2_{s, 1}$ therefore corresponds to the most likely characteristic strain value, i.e., the peak of $p(h^2_c)$, which can be predicted as
\begin{equation}
    h^2_{c, {\rm peak}} = \int_0^{h^2_{s, 1}} dh^2_s \frac{dN}{dh^2_s} h^2_s.
    \label{eq:h2peak}
\end{equation}
The peak of the distribution is therefore produced by the large number of faint sources with brightness below $h^2_{s, 1}$, and this estimate is identical to the estimate from Ref.~\cite{Sesana:2008mz} when integrating over the chirp mass.

\begin{figure*}[t]
    \centering
    \includegraphics[width=0.8\textwidth]{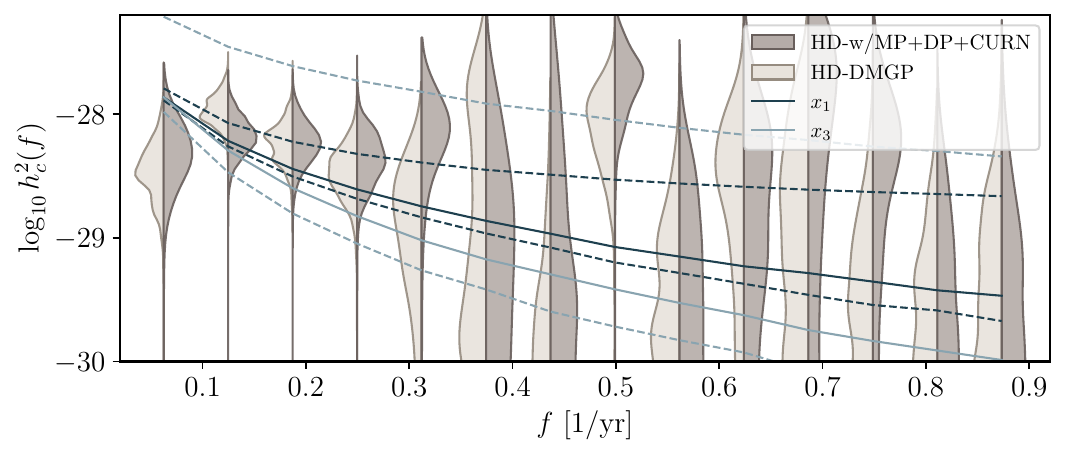}
    \caption{Posterior distributions of the characteristic strain squared for the first 14 frequencies in the 15-yr NANOGrav~\cite{NANOGrav_stoc, Lamb_freespec} analysis. The dark grey violin plots on the right correspond to the HD correlated timing residuals, where a monopolar, dipolar, and uncorrelated red noise components are simultaneously fit, while the lighter violins on the left result from an equivalent analysis using an alternative dispersion measure model. The solid lines labelled $x_1$ and $x_3$ show examples of the median GWB for models that fit the observed spectra within 1- and 3-$\sigma$ confidence intervals. The dashed lines show the values of the 5 and 95\% bounds of the model distribution $p(\log_{10} h^2_c|\vec{\theta})$.}
    \label{fig:h2c_free-spec}
\end{figure*}

\subsection{Merger history}
In the preceding discussion, we assumed a scenario in which only the most recent merger significantly contributes to the SGWB. Here we explore the effect of multiple equal mass mergers on the characteristic strain distribution function as an upper limit of the SGWB prediction. We recall that, as in Ref.~\cite{Sato-Polito:2023gym}, the impact of multiple mergers on the mean characteristic strain squared is, at most, an increase by a factor of $~2.7$ relative to the single merger scenario.

As in Ref.~\cite{Sato-Polito:2023gym}, we account for the redshift distribution of each previous merger $p_{N+1}$ by fixing the redshift distribution of the latest merger $p_{N=1}$, given in Eq.~\ref{eq:pz_pq}, and choosing a fixed time delay $\tau$. Each redshift $z_{N}$ is then mapped onto the redshift of the previous merger via $z_{N+1} = z_{N}+\Delta z(\tau)$. Since all mergers are assumed to be between equal mass partners, the mass function of a previous merger can be found simply by
\begin{equation}
    \phi_N(M) = 2^{N-1} \phi_{1}(2^{N-1} M),
\end{equation}
where $\phi_{1}$ is given by Eq.~\ref{eq:MF_scatter}. At each previous merger, the luminosity function is boosted at the faint end and suppressed at the bright tail, as shown in Fig.~\ref{fig:mu-multiple-gens} in Appendix \ref{app:dist_power_realistic}. In general, the effect of including multiple mergers on the characteristic strain distribution function is an increased contribution from a large number of faint sources. This increases the amplitude of the background and the characteristic number of sources, while decreasing the peak mass and strain amplitude that most contributes. A quantitative discussion of the changes to $p(h^2_c)$ in the case of multiple instantaneous mergers is given in App.~\ref{app:multiple-gen-mergers}.

The total luminosity function is therefore the sum of the luminosity function of each layer. In order to compute the characteristic strain distribution in the scenario with multiple mergers, we therefore simply modify the luminosity function and follow the same procedure outline in Sec.~\ref{sec:hc_pdf}. 

We emphasize that we have implicitly assumed that black holes do not accrete at all, which leads to an extremely conservative estimate --- since our prediction is anchored on the present-day black hole mass function, including accretion would greatly diminish the contribution from previous mergers. To show this, suppose that early in its inspiral (before entering the PTA band), every binary had an initial mass that is a fraction $\kappa $ of its final mass. That is, if in the latest merger $N=1$ a binary had a total mass $M$, we assume that it accreted a mass $M_{\rm acc} = M(1-\kappa )$ since its previous merger, such that in the previous layer $N=2$ we now have twice as many binaries binaries each with total mass of $M'=\kappa M/2$. In this scenario, we now have that the series for the characteristic strain is given by
\begin{equation}
    h^2_{c, N} = h^2_{c, 1}\left( \sum_{N=1}^{\infty} 2^{-2(N-1)/3} \kappa ^{5(N-1)/3} \right),
    \label{eq:h2c_acc}
\end{equation}
where we have neglected the possibility of accretion between the latest merger and the single black hole observed today at $z=0$. Hence, any amount of accreted mass will diminish the series by an additional factor of $\kappa ^{5/3}$. For instance, if the accreted mass is comparable to the initial mass, say $\kappa =1/2$, leads to $h^2_{c, N} = 1.25 h^2_{c, 1}$. Hence, while there are proposed mechanisms that preferentially lead to equal mass mergers due to differential accretion~\cite{2024arXiv240117355V, 2020ApJ...901...25D, 2020MNRAS.498..537S}, it is unlikely that the boost from multiple mergers would be significant in this scenario. We also note that the most massive SMBHs are hosted by massive elliptical galaxies with low-density centers. The leading explanation for core formation is stellar scattering with the SMBHs after gas-poor mergers \cite{1997AJ....114.1771F, 2001ApJ...563...34M, 2024MNRAS.528....1H}, also disfavoring a scenario in which accretion plays a significant role. In summary, the agreement between the PTA measurement and local scaling relations requires a physical mechanism in which SMBHs merge extremely quickly, while minimizing accretion, which is yet unknown. 

\section{SGWB spectrum}\label{sec:PTA_spectrum}

For each pulsar monitored by a PTA, a deterministic timing model is constructed, which includes a variety of terms that affect the arrival time of pulses. These include the spin period and its derivative, the effect of a companion if the pulsar is in a binary system, the pulse propagation in the interstellar medium, astrometric effects such as parallax, proper motions, among others. The timing residuals $\vec{\delta t}$ are then defined as the difference between the observed arrival times and the deterministic model, which captures what are assumed to be red or white stochastic processes. Evidence for a GWB is then expected to arise as the variance of a red-noise process correlated among pulsars.

We will compare the models for the GWB discussed in Sec.~\ref{sec:hc} with the spectrum measured by the NANOGrav~\cite{NANOGrav_stoc} collaboration. We use the free-spectrum posterior \cite{Lamb_freespec} of the cross-correlated timing-residual power spectral density (PSD), which is related to the characteristic strain in each frequency bin centered in $f_i$ by
\begin{equation}
    \rho^2(f_i) = \frac{h^2_c(f_i)}{12\pi^2 f_i^3 T_{\rm obs}}.
\end{equation}
Given a set of model parameters $\vec{\theta}$, the PTA likelihood can then be written as
\begin{equation}
    p(\vec{\delta t}|\vec{\theta}) \propto \prod_{i=1}^{N_f} \int d \rho_i \frac{p(\rho_i|\vec{\delta t})}{p(\rho_i)} p(\rho_i|\vec{\theta}),
\end{equation}
where $p(\rho_i|\vec{\delta t})$ are the posterior probability distributions of the PSD, $p(\rho_i)$ is the prior distribution, $p(\rho_i|\vec{\theta})$ is the probability of observing a GWB given a set of model parameters, and $N_f$ is the number of frequency bins used in the analysis. Note that each frequency bin is assumed to be independent, which was investigated in Ref.~\cite{Lamb_freespec}.

Similarly to Ref.~\cite{NANOGrav_SMBH}, we choose the free-spectrum posteriors of the Hellings-Downs correlated signal, in which a monopole- and dipole-correlated (MP and DP) red noise, and a common uncorrelated red noise (CURN) are simultaneously modelled, referred to as HD-w/MP+DP+CURN. An alternative analysis is performed with a different model for the dispersion measure (DM), labelled HD-DMGP. The latter yields a slightly larger amplitude for the GWB and a different spectral shape. As discussed in Ref.~\cite{NANOGrav_stoc}, only the first 14 frequency bins display evidence for a common red noise spectrum. The Bayes factor between the model for a HD-correlated red noise and a CURN, both described by a power law with free spectral index, drops from 1000 to 200 when frequencies beyond the first 5 are included. However, this is potentially due to the lack of power observed in the 6th and 7th frequency bins, while an ideal power law spectrum entails a monotonic spectrum. While this motivated Ref.~\cite{NANOGrav_SMBH} to focus only on the first 5 bins (finding similar results if all 14 were used), we choose to include all 14 bins in our fiducial analysis, since an astrophysical GWB may indeed fluctuate frequency to frequency. Our main results thus focus on the HD-w/MP+DP+CURN posterior with 14 frequency bins.

Fig.~\ref{fig:h2c_free-spec} shows the free-spectrum distributions of the characteristic strain squared for the two different analyses described above (HD-w/MP+DP+CURN and HD-DMGP). The lines correspond to an example of an astrophysical model consistent with the data within the 1- and 3-$\sigma$ confidence intervals, where the median of the $h^2_c$ model distribution shown in the solid line and the 5 and 95\% limits of the distribution in dashed. From this example, it is evident that the inclusion particularly of the 5th and 8th frequency bin (the only bin above the first 5 with significant evidence for correlated power) will lead to a preference for SMBH models dominated by a larger number of sources. We also note that the only physical mechanism for a reduction in power in the lowest frequencies are interactions between binaries and their environment, which we do not include in our model. The low power seen in the first frequency bin will therefore never be well described, potentially resulting in a slightly lower amplitude fit.

\section{Results}\label{sec:results}
\subsection{Isotropic Background}

\begin{figure}
    \centering
    \includegraphics[width=0.45\textwidth]{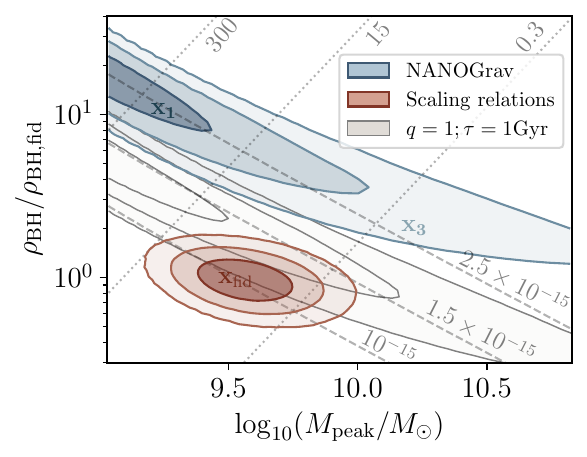}
    \caption{Posterior distributions for the peak mass contributing to the SGWB $M_{\rm peak}$ and the black hole mass density $\rho_{\rm BH}$ relative to its fiducial value of $\rho_{\rm BH, fid}=4.5\times 10^5 M_{\odot}$Mpc$^{-3}$. The NANOGrav results using the 15 yr Hellings-Downs correlated free spectrum, with a monopole- and dipole-correlated red noise, and common uncorrelated red noise (HD-w/MP+DP+CURN) modelled simultaneously are shown in blue, while the red contours show the values inferred from the present-day black hole mass function. The three contours show the 1-, 2-, and 3-$\sigma$ regions. The grey contour corresponds to an upper limit scenario, in which all black holes are assumed to merge multiple times with equal mass partners. Results in both grey and blue make the optimistic assumption of neglecting accretion, which would only diminish the GW signal. The dashed grey lines show lines of constant characteric strain, while the dotted lines correspond to constant characteristic number of sources.}
    \label{fig:N_Mpeak}
\end{figure}

While in principle the SGWB depends on a variety of properties SMBHs, and the redshift and mass ratio distributions of mergers, which we specified through numerous parameters in Sec.~\ref{sec:mean_h2c}, these quantities are highly degenerate. Given these parameter degeneracies and the relatively large uncertainties on the current GW measurement, it is both convenient and accurate to summarize the SGWB model in terms of a smaller subset of parameters that are sufficient to describe the data. Assuming a paradigm in which the SGWB is produced by astrophysical sources, the measured background can be very well described in terms of an overall amplitude and a spectral dependence that may deviate from the expected mean ($h^2_c \propto f^{-4/3}$) due to Poisson fluctuations in the number of sources, which is captured by the characteristic number of sources $N_c$. Due to the approximate scalings presented in Eqs.~\ref{eq:h2c_scaling} and \ref{eq:Nc_h2c_scaling}, the physical quantities constrained by the PTA data correspond to an overall amplitude, which we express in terms of the black hole mass density $\rho_{\rm BH}$, and a peak mass contributing to the GW signal $M_{\rm peak}$. In practice, the black hole mass density is effectively varied through the amplitude of the SMBH mass function, while $M_{\rm peak}$ is varied through the scatter in the $M-\sigma$ relation. 

Fig.~\ref{fig:N_Mpeak} shows the posterior distribution of $\rho_{\rm BH}$ and $M_{\rm peak}$ by fitting the NANOGrav 15 yr free spectrum. The red contour shows the values of $\rho_{\rm BH}$ and $M_{\rm peak}$ implied by estimates of the present-day black hole mass function, and is obtained by assuming Gaussian errors on the parameters that describe the velocity dispersion function and the scaling relation, with values given in Sec.~\ref{sec:mean_h2c}. The dashed lines show lines of constant $h^2_c$ amplitude, while the values to the right are $h_c(f=$yr$^{-1})$. The dotted lines correspond to constant $N_c$ and the values at the top show the particular values for the fifth frequency bin ($f=0.3$yr$^{-1}$). The shape of the NANOGrav contour can be understood from the fact that, at present, the strain amplitude is the main quantity that can be measured and therefore the posterior mostly follows the dashed lines. The slight spectral information leads to the shift towards low $M_{\rm peak}$ and higher amplitudes. 

We highlight that the PTA fit shown in Fig.~\ref{fig:N_Mpeak} is generally consistent with the NANOGrav analysis~\cite{NANOGrav_SMBH}. We differ, however, in the interpretation of these results, primarily due to the explicit comparison to the present-day SMBH mass function. Indeed, Ref.~\cite{NANOGrav_SMBH} also finds a preference for models in which the background amplitude is high (captured in the high amplitude of the galaxy stellar mass function), in which many sources contribute (high stellar mass function and low scatter in the scaling relation), and with the lowest possible time delay between mergers (peaked at $\tau=0$).

The results shown in Fig.~\ref{fig:N_Mpeak} are also consistent with earlier work from Ref.~\cite{Sato-Polito:2023gym}, in which we noted that a $\rho_{\rm BH}$ between $4-10$ times larger was required to fit the characteristic strain spectrum at 90\% CL for a value of $M_{\rm peak}$ consistent with inferences of the present day SMBH mass function based on scaling relations and galaxy velocity or stellar mass functions. Since in Ref.~\cite{Sato-Polito:2023gym} we only considered the amplitude of a power law with slope fixed to the mean expectation from SMBHBs and compared it to the mean characteristic strain amplitude, Fig.~\ref{fig:N_Mpeak} suggests a slightly higher value due to the difference in the median of $p(h^2_c)$ and the mean of the distribution, as discussed in Sec.~\ref{sec:hc_pdf}.

The grey contour in Fig.~\ref{fig:N_Mpeak} shows the result assuming all mergers occur between equal mass partners, with an extremely short time delay of $\tau=1$Gyr, and assuming that no mass is accreted onto the black holes. This results in an upper limit scenario in which PTA measurements and local scaling relations are more consistent, and is equivalent to the upper limit discussed in Ref.~\cite{Sato-Polito:2023gym}.

\begin{figure}
    \centering
    \includegraphics[width=0.45\textwidth]{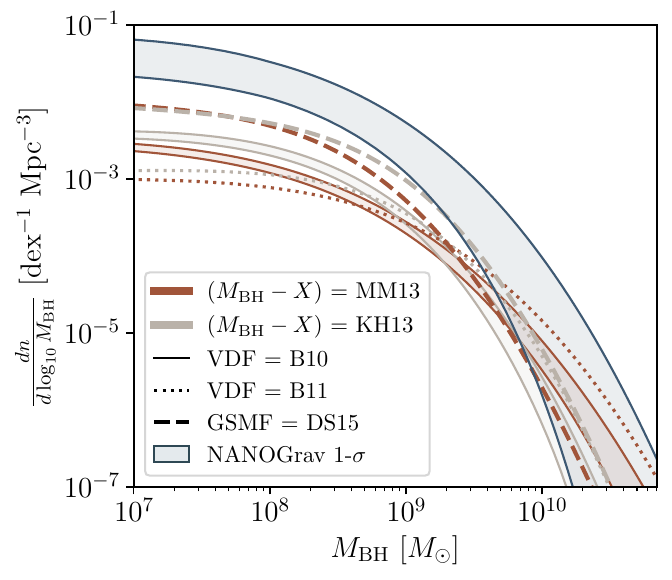}
    \caption{Comparison between the single black hole mass function inferred from a combination of scaling relations and galaxy catalogs with the mass function implied by the NANOGrav 15-yr results. The dark blue shaded region corresponds to the 1-$\sigma$ contour of Fig.~\ref{fig:N_Mpeak}. The scaling relations (labelled MM13 and KH13) are presented in \cite{mcconnell_ma} and Refs.~\cite{kormendy_ho}, respectively, while the velocity dispersion functions (B10 and B11) are measured in Refs.~\cite{2010MNRAS.404.2087B} and \cite{2011ApJ...737L..31B} and the GSMF is given in \cite{2015MNRAS.454.4027D} (DS15). Similarly to Ref.~\cite{Sato-Polito:2023gym}, we have omitted the error bands of most SMBH mass functions for visual clarity, showing only for the B10 VDF result since it corresponds to the fiducial model adopted in this work.}
    \label{fig:BHMF}
\end{figure}

Finally, we show in Fig.~\ref{fig:BHMF} a direct comparison between the remnant single black hole mass functions that fit the NANOGrav 15-yr spectrum in Fig.~\ref{fig:N_Mpeak} and mass functions directly inferred from a combination of scaling relations and galaxy catalogs. The scaling relations are measured in Refs.~\cite{mcconnell_ma} and Refs.~\cite{kormendy_ho}, while the galaxy samples are presented in Refs.~\cite{2010MNRAS.404.2087B}, \cite{2011ApJ...737L..31B}, and \cite{2015MNRAS.454.4027D}. These choices of scaling relations and the galaxy catalogs are discussed in greater detail in Ref.~\cite{Sato-Polito:2023gym}, but are adopted here as representative examples of black hole mass functions. All estimates based on scaling relations are consistent with each other, while underpredicting the GWB signal (see discussion in Ref.~\cite{Sato-Polito:2023gym}).

\subsection{Individual Sources}

\begin{figure*}[t]
    \centering
    \includegraphics[width=0.8\textwidth]{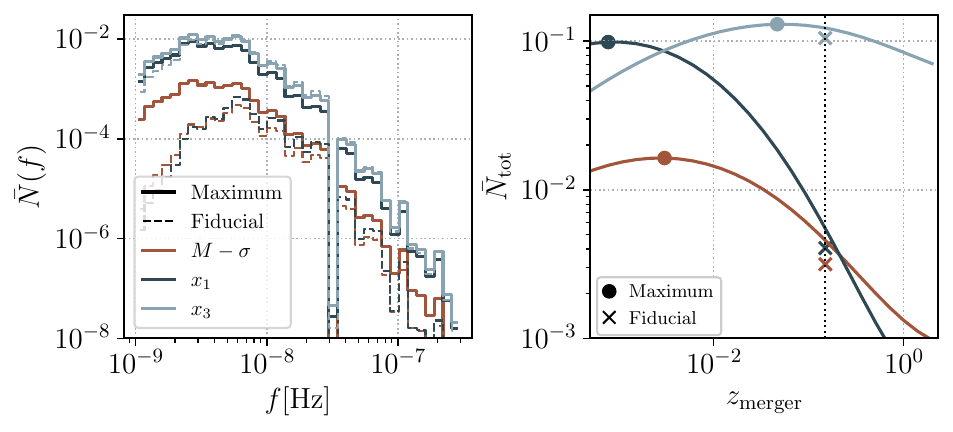}
    \caption{Mean number of detectable sources given the current sensitivity of the individual source search from NANOGrav~\cite{NANOGrav_individual}, for different models consistent with the isotropic background. The curves labelled $M-\sigma$, $x_1$, and $x_3$ correspond to mass functions consistent with the local mass function estimate and points along the degeneracy of the contour in Fig.~\ref{fig:N_Mpeak}, with $(\rho/\rho_{\rm fid}, \log_{10} M_{\rm peak}) = (1, 9.5), (10, 9.25), (2, 10.3)$, respectively. The points $x_1$, and $x_3$ are within the 1- and 3-$\sigma$ confidence levels of the NANOGrav posterior. The solid curves on the left and dots on the right correspond to the optimal redshift distribution that leads to the maximum number of detectable sources, while the dashed (left) and crosses (right) show the values for the fiducial redshift distribution given in Eq.~\ref{eq:pz_pq} and adopted throughout this work. The panel on the right shows the total number of detectable sources (across all frequencies) assuming that all black holes merge at a single redshift $z_{\rm merge}$, while the dotted vertical line shows the redshift from which most of the characteristic strain signal is sourced in the fiducial redshift distribution. Notice that the typical redshift for which the maximum number of detectable sources is achieved is unphysically small and only corresponds to a mathematical upper limit.}
    \label{fig:single_source}
\end{figure*}

One natural question that arises is whether the region of parameter space in Fig.~\ref{fig:N_Mpeak} found to fit the stochastic background is consistent with the non-detection of individual sources presented in Ref.~\cite{NANOGrav_individual}. It is possible that the high $M_{\rm peak}$ tail would lead to a significant number of binaries that would be individually detectable and could therefore be ruled out. In summary, we ask: can the SMBH distribution be better constrained by considering the non-detection of point sources?

The individual source search in PTA data describes a single SMBH binary as a continuous wave and searches the timing residuals of all pulsars for this deterministic signal. A non-detection can then be translated into an upper limit on the GW amplitude of a source as a function of sky position and frequency. Given an upper limit on the strain amplitude $h_{\rm ul}$, where the inclination and polarization averaged value is defined as
\begin{equation}
    h=\frac{8}{\sqrt{10}} \left(\frac{G M}{c^2}\right)^{5/3} \eta \frac{(1+z)^{2/3}}{\chi(z)} \left(\frac{\pi f}{c}\right)^{2/3},
    \label{eq:h0}
\end{equation}
the expected number of detectable sources per logarithmic frequency is given by 
\begin{equation}
    \frac{d\bar{N}}{d\log f} = \int^{\infty}_{h_{\rm ul}} dh \frac{dN}{dh d\log f},
    \label{eq:Nlogf_det}
\end{equation}
which can be computed similarly to Eq.~\ref{eq:dNdhsdlogf} and we use the upper limit $h_{\rm ul}$ presented in Ref.~\cite{NANOGrav_individual}.

The number of detectable sources in a given frequency bin therefore depends on the high $h$ tail of the luminosity function. Models with a higher peak mass (or a correspondingly high amplitude), or in which the unknown redshift distribution of sources is dominated by low redshifts will typically have a larger number of detectable sources. In particular, since the brightness of a source scales with $1/\chi$, the choice of redshift distribution will be especially impactful. We note, however, that due to the fact that a redshift distribution peaked at lower $z$ results in a contribution from a smaller volume, there is always an optimal redshift distribution for which the number of detectable sources at a given frequency is maximal. To see this, consider the low-mass/low-strain limit of the SMBH mass function, where $dn/d\log M$ is approximately constant. In this limit, we can show by combining Eqs.~\ref{eq:dlogfdt}, \ref{eq:dVdt}, and \ref{eq:luminosity-function-Mc} that the kernel of the redshift integral is given by
\begin{equation}
    \frac{dN}{dh d\log f} \propto \int dz\ p_z(z) \frac{\chi(z)}{1+z},
\end{equation}
which peaks at $z \sim 1.6$ and the optimal $p_z(z)$ is therefore a delta function centered on this peak redshift. For larger masses, which correspond to higher strain amplitudes or lower frequencies, the black hole mass function is steeper than in the low mass limit, say $dn/d\log M \propto M^{-\alpha}$. Then, from Eq.~\ref{eq:h0}, we may conclude that the redshift kernel above receives an additional contribution of $\left[\chi^{3/5}/(1+z)^{2/5}\right]^{-\alpha}$, which pushes the peak towards lower values. Hence, there is always a redshift distribution for which the number of detectable sources is maximized, given a sensitivity to the GW amplitude as a function of frequency. 

We show in Fig.~\ref{fig:single_source} the mean number of detectable sources for different assumptions regarding the SMBH mass function and redshift distribution. The lines labelled $x_1$ and $x_3$ correspond to parameter values consistent within the 1- and 3-$\sigma$ confidence levels shown in Fig.~\ref{fig:N_Mpeak}, chosen along the degeneracy axis, while the line labelled $M-\sigma$ is consistent with the present-day black hole mass function estimate. Given a particular SMBHB distribution, we compute the number of detectable sources from the upper limits on the individual source strain amplitude placed by Ref.~\cite{NANOGrav_individual} using Eq.~\ref{eq:Nlogf_det}, where $\bar{N}(f) = \frac{d \bar{N}}{d \log f} \Delta \log f$ and $\bar{N}_{\rm tot}$ is the sum over all frequencies. The right panel of Fig.~\ref{fig:single_source} shows $\bar{N}_{\rm tot}$ under the assumption that all black holes merge at a single redshift $z_{\rm merge}$. The redshift distribution that leads to the highest number of detectable sources is, in most cases, unphysically low and the peak value of $\bar{N}_{\rm tot}$ should therefore be interpreted as a mathematical upper limit. The fiducial value for $\bar{N}_{\rm tot}$ marked in crosses are slightly lower than the solid curves due to the fiducial distribution being extended in redshift, while the solid line assumes a single merger redshift.

We conclude that all SMBH distributions consistent with the SGWB measurement are also consistent with the non-detection of an individual source, since the expected number of detectable sources is, at most, $\lesssim 0.1$. While the results shown in this section focus on a scenario in which only the latest merger significantly contributes, multiple merger events will only increase the number of fainter sources, and would therefore lead to a smaller number of detectable sources (for the same background amplitude). If indeed the discrepancy pointed out in this work is solved by a combination of multiple merger events, nearly equal mass mergers or a high GW measurement, Fig.~\ref{fig:single_source} suggests that the mean number of detectable sources for the current PTA sensitivity is at most $\bar{N}\lesssim 0.01$ and is therefore unlikely to be detectable in the near future. An improvement in sensitivity of around a factor of $\sim 10$ would be required for $\bar{N}_{\rm tot}$ to be $\sim 1$ for a mass function consistent with local scaling relations and galaxy surveys and the fiducial redshift distribution. 

\section{Conclusion}\label{sec:conclusion}
The recent discovery of a nano-hertz gravitational-wave background raises a variety of questions regarding the origin of the signal and the nature of its sources. The most likely candidate is a population of inspiralling supermassive black holes, raising the natural question of what we can learn regarding the source population and whether it is consistent with predictions based on electromagnetic observations. Similarly to Ref.~\cite{Sato-Polito:2023gym}, we compare the GW measurement made by PTAs with estimates of the background based on the remnant population of single black holes today \cite{Phinney:2001di}. We complement the previous analysis by using the full spectral information captured by the PTA data set.

Under the assumption that the background is produced by SMBHs, a discrete number of point sources contribute to the background at each frequency, which can lead to deviations in the observed spectrum relative to the average ($h^2_c \propto f^{-4/3}$) due to Poisson fluctuations \cite{Sesana:2008mz}. This is one of the key features that can offer additional information about the source population beyond the amplitude of the spectrum. To account for this potential spectral deviation, we derive a simple analytical expression for the distribution function of characteristic strains $p(h^2_c|\vec{\theta})$, given a set of SMBH parameters $\vec{\theta}$. 

We refit the free-spectrum posterior of the pulsar timing residuals from NANOGrav's 15-yr data set using a minimal model for the characteristic strain as a function of supermassive black hole properties described above. We conclude that: (1) similarly to Ref.~\cite{Sato-Polito:2023gym}, around $5-12$ times more black holes (and mass density) are required to match estimates based on the present-day black hole mass function, and (2) the shape of the GWB spectrum already provides additional information to the amplitude, indicating a preference for a SMBH mass function that is dominated by a large number of fainter sources (the typical mass that contributes to the background is $\lesssim 10^{10} M_{\odot}$ within a $2\sigma$ confidence interval).

Finally, we show that the region of parameter space consistent with the isotropic background measurement is entirely consistent with the non-detection of individual sources \cite{NANOGrav_individual}. While the prediction of the mean number of detectable sources for a given sensitivity depends on the uncertain redshift distribution of SMBHBs, we show that an upper limit can always be derived, albeit typically requiring black holes to merge at extremely low redshifts. In the fiducial scenario, the mean number of detectable sources is typically a few times $\sim 10^{-3}$ for most models consistent with the isotropic background. At the extreme end of SMBH mass functions dominated by few very massive sources, but still consistent with the isotropic background within 3-$\sigma$, a mean number of detectable sources around $\bar{N} \lesssim 0.1$ could be expected. 

\acknowledgments
We would like to thank Eliot Quataert for helpful discussions and suggestions. GSP gratefully acknowledges support from the Friends of the Institute for Advanced Study Fund. MZ is supported by NSF 2209991 and NSF-BSF 2207583.

\bibliography{ref.bib}
\bibliographystyle{utcaps}

\onecolumngrid
\appendix
\section{Distribution of power for power law luminosity functions}\label{app:dist_power}

In this appendix we derive the distribution for the total power under the assumption that the luminosity function of sources is a power law:
\[
\frac{dN}{dh_s^2} = \frac{N_p}{h_{p}^2} (h_{s}^2/h_{p}^2)^{-\gamma},
\]
where we have defined a pivot point and $N_p$ is the number of sources per logarithmic interval at the pivot point and $h_{p}^2$ is the square of the strain of the sources at the pivot point. This distribution is assumed to hold for a limited range in $h^2_s$ roughly centered at the pivot point. We will be interested in $1 < \gamma < 3$. 

This simplifying assumption will allow us to get some analytic results to build some intuition on the general shape of the distribution and the properties of its tail. We will conclude that:
\begin{itemize}
    \item The distribution looks like a Gaussian-like core and a power law tail.
    \item The power law tail is directly given by the luminosity function of sources, the easiest way to obtain a very high value of the total power is to have one very bright source.
    \item   The Gaussian-like core is determined by the numerous and faint sources. As their importance grows as the shape of the distribution is progressively closer to a Gaussian. 
\end{itemize}

There is of course freedom in choosing $N_p$ and $h_{p}^2$ in our power law distribution. An important consideration is that in the faint end of the power law, the sources are very numerous and thus the central limit theorem ensures that their contribution to the $h_{t}^2$ is distributed as a Gaussian. The interesting part of the distribution is related to the sources that are not too numerous. Based on this we will pick the pivot point at the place such that the expected number of sources above it is one. Furthermore we can measure the strain in units of $h_{p}^2$ by defining:
\[
\frac{dN}{dx} = (\gamma -1) x^{-\gamma} \ \ ; \ \ x=\frac{h_{s}^2}{h_{p}^2}.
\]
We will assume that this distribution is valid over a range $x_{\rm min}<x<x_{\rm max}$ with $x_{\rm min}<<1$ and $x_{\rm max}>>1$. The total power from sources below $x_{\rm min}$ is distributed as a Gaussian to a very good approximation, while we will choose $x_{\rm max}$ such that the sources above it are too rare to make any difference. For the numerical results presented below we will pick $x_{\rm min}= 1/x_{\rm max}=10^{-2}$.  

For this distribution:
\begin{align}
{\bar N} =& \int_{x_{\rm min}}^{x_{\rm max}} \frac{dN}{dx}  dx = (x_{\rm min}^{1-\gamma}-x_{\rm max}^{1-\gamma}) \nonumber \\
    N(x>1) = & \int_1^{\infty} \frac{dN}{dx} dx = 1 \nonumber \\
    {\bar x_t} =&  \int_{x_{\rm min}}^{x_{\rm max}} \frac{dN}{dx} x \  dx = \frac{\gamma-1}{2-\gamma} (x_{\rm max}^{2-\gamma}-x_{\rm min}^{2-\gamma}) \nonumber \\
    {\rm var}(x_t) = & \int_{x_{\rm min}}^{x_{\rm max}} \frac{dN}{dx} x^2 \  dx = \frac{\gamma-1}{3-\gamma} (x_{\rm max}^{3-\gamma}-x_{\rm min}^{3-\gamma}) 
\end{align}

The slope $\gamma=2$ is special, it corresponds to the case when there is equal contributions to $\bar{x}_t$ per logarithmic bin in $x$. For $\gamma < 2$ bright sources dominate and for $\gamma > 2$ faint sources dominate. In all cases the actual value of $\bar{x}_t$ depends on the cut-off of the integral, this is why it for a power law it is useful to compute the distribution of fluctuations around the mean. For the entire range of interest, $1 < \gamma < 3$ the variance is dominated by the contribution from the bright sources while the total number of sources is dominated by the contribution from the faint ones. 

For this distribution we can calculate $u(\omega)$:
\begin{equation}
    u(\omega) = (\gamma-1) \Gamma(1-\gamma) \omega^{\gamma-1} e^{i (\gamma-1) \pi/2},
\label{eq:u_h2}
\end{equation}

and use it to compute the distribution $P_\gamma(x_t)$ of the total power $x_t$. 

\begin{figure}
    \centering
    \includegraphics[width=0.85\linewidth]{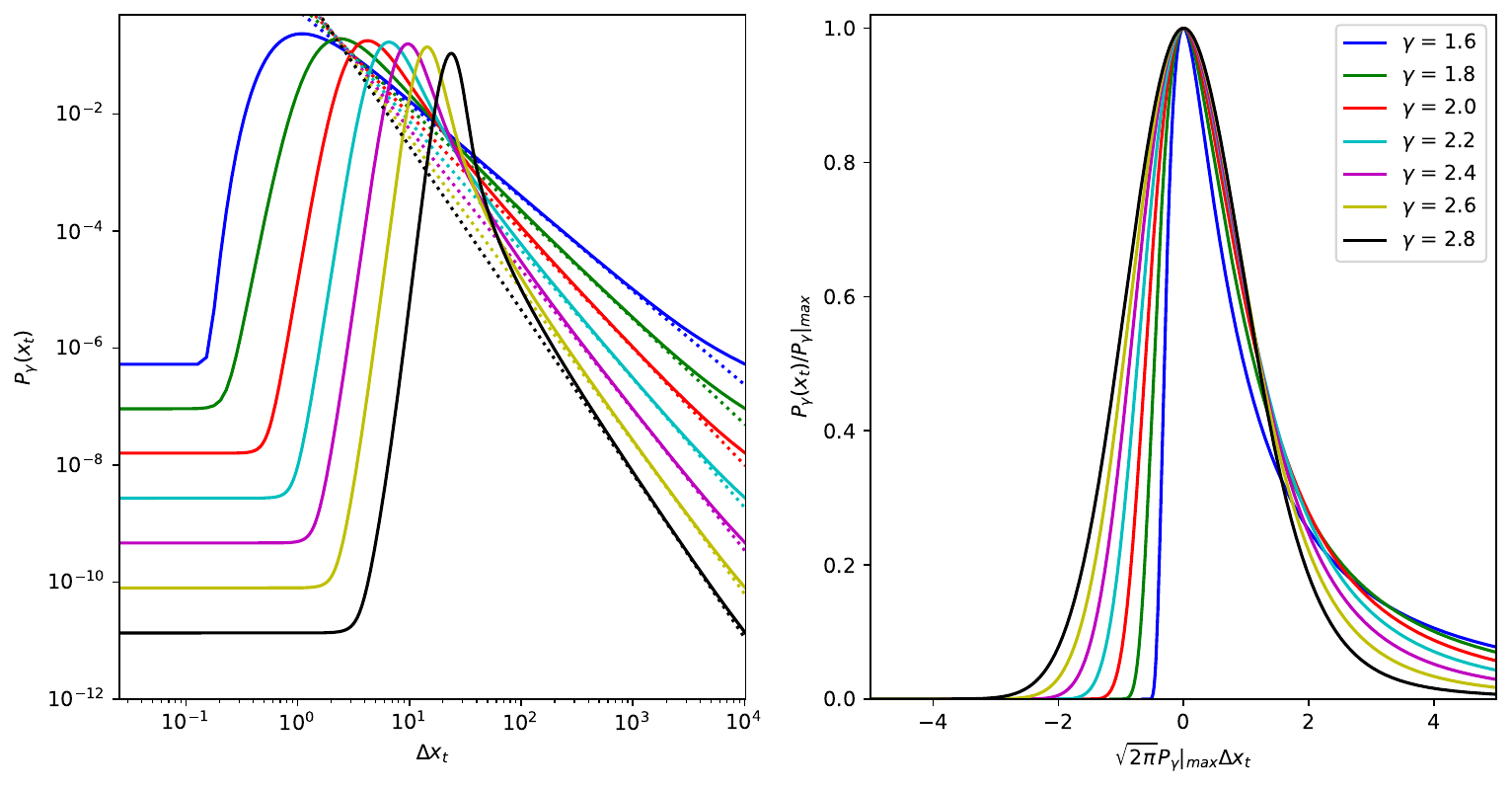}
    \caption{$P_\gamma$ for several values of $\gamma$. The curves were obtained by doing a fast Fourier transform of the analytic result based on \ref{eq:u_h2}. The chosen range of $x$ was $10^{-2} < x < 10^{2}$.   Artifacts can be seen due to the assumed periodicity in the Fourier transform when one gets to the boundary (left panel). The curves have been shifted by a different amount in each panel for clarity. In the left panel the minimum probability was set at the origin. In the left panel, in addition to $P_\gamma$ we also show $dN/dx$. It is clear that to the right of the Gaussian-like core, when the probability becomes low, the tail of the distribution is given by $dN/dx$. The departures at very high $x$ are an artifact of the periodic boundary conditions.  On the right the peaks have been aligned and normalized to one. If the distributions were Gaussian, the standard deviation of the Gaussian would be given by $\sigma=1/\sqrt{2\pi}P_\gamma|_{\rm max}$. We used this quantity to re-scale the $x$ axis. In the right panel one can see how the Gaussian-like core accounts for more of the distribution. As $\gamma$ changes the width of the Gaussian-like core changes, the distribution is narrower for smaller $\gamma$s (the width for $\gamma=2.8$ is almost 4 times larger than for $\gamma=1.6$). The mean evolves dramatically and is dominated by the tail. As result it becomes larger for smaller values of $\gamma$. In fact for the distribution where the peaks have been aligned, the mean is almost eighty for $\gamma=1.6$ but about one for $\gamma=2.8$.}
    \label{fig:Ggamma}
\end{figure}

There are several properties of $P_\gamma$ that are worth stressing. The general form of $P_\gamma$ is that of  a Gaussian-like core and a tail. Figure \ref{fig:Ggamma} illustrates the fact that the tail is directly given by ${dN}/{dx}$, 
\[
P_\gamma(x_t) \sim \frac{dN}{dx}\ \  (x_t \gg 1)
\]
reflecting the fact that at high $x_t$ the probability is dominated by the chances of having one source of with $x \sim x_t$. 

The properties of the Gaussian-like core however are determined by the numerous faint sources. As an estimate of the location and width of the Gaussian core one can compute the contribution to the mean and variance coming from sources with $x<1$. This is just an estimate as the transition between numerous and rare sources is somewhat arbitrary.  However this simple estimate captures the main features of the results well. For sources with $x<1$ we have:
\begin{align}
    {\bar x_t}|_{x<1} =&  \int_{x_{\rm min}}^{1} \frac{dN}{dx} x \  dx = \frac{\gamma-1}{2-\gamma} (1-x_{\rm min}^{2-\gamma}) \nonumber \\
    {\rm var}(x_t)|_{x<1} = & \int_{x_{\rm min}}^{1} \frac{dN}{dx} x^2 \  dx = \frac{\gamma-1}{3-\gamma} (1-x_{\rm min}^{3-\gamma}). 
\end{align}
Both the mean and the variance increase with $\gamma$ over the range of interest as shown in \ref{fig:Ggamma}. 

The figure also illustrates the fact that the location of the Gaussian bump is a much stronger  function of $\gamma$ than the mean. This is so because in the $\gamma$ range of interest one goes through the special $\gamma=2$ case and thus the answer is also sensitive to the value of $x_{\rm min}$ which dominates the integral for $\gamma>2$. One should keep in mind however that the sources with $x\ll 1$ are very numerous and thus the central limit theorem guarantees that the resulting distribution for $x_t$ that they produce would be very well approximated by a Gaussian. Thus as $\gamma$ increases the contribution from the faint sources increases and thus the Gaussian bump moves to the right to account for them, while the shape of $P_\gamma$ becomes more and more Gaussian. 

The behavior of the width of the Gaussian bump is quite different. In all the range of interest, it is the $x=1$ limit of the integral that dominates. Thus the width is fairly insensitive to $\gamma$ and only starts to increase more rapidly when $\gamma \rightarrow 3$ which is the point where the variance would receive equal contributions from each logarithmic  range of $x$. 

The results in this section provide an intuition of what on should get for a realistic luminosity function. In fact for realistic models the luminosity function, it is well approximated by a power law with a slope that  changes slowly. 

\section{Distribution of power: realistic case}\label{app:dist_power_realistic}

In the main text we presented results for the distribution of power for realistic luminosity functions for the sources. In this appendix we present some additional details that will help build some intuition to asses the robustness of our results. In addition we report the details needed for an accurate numerical calculation of the distribution of power. 

It is convenient to change variables to the characteristic number of sources $N_c$ defined in Eq.~\ref{eq:Nc_def} and the characteristic source power $x$: 
\[
x=\frac{h_s^2}{h_s^2|_{\rm peak}}.
\]
These variables are convenient because they absorb most of the parameter dependence.  In terms of these quantities we can write the luminosity function of sources as:
\[
\frac{dN}{d\log x} = N_c \mu(x),
\]
where $\mu(x)$ is a somewhat universal function in the sense that it depends weakly on the shape of the mass function of black holes.  The dependence on the frequency is implicit through the definition of $x$ and $N_c$.  Thus rather than having to compute the distribution as a function of $\phi_\star$, $M_{\rm peak}$ and $f$ one only needs to do a computation as a function of one parameter $N_c$. In the main sections we change $M_{\rm peak}$ by changing the scatter in the $M-\sigma$ relation rather than the characteristic velocity dispersion in the velocity function of galaxies. In that case the shape of $\mu(x)$ evolves slowly with the parameters. This is included in our results but is a small correction. In this appendix we will consider a single universal form for $\mu(x)$.

With these definitions $\mu(x)$ satisfies:
\[
\int \mu(x) dx = 1.
\]

\begin{figure}
    \centering
    \includegraphics[width=0.85\linewidth]{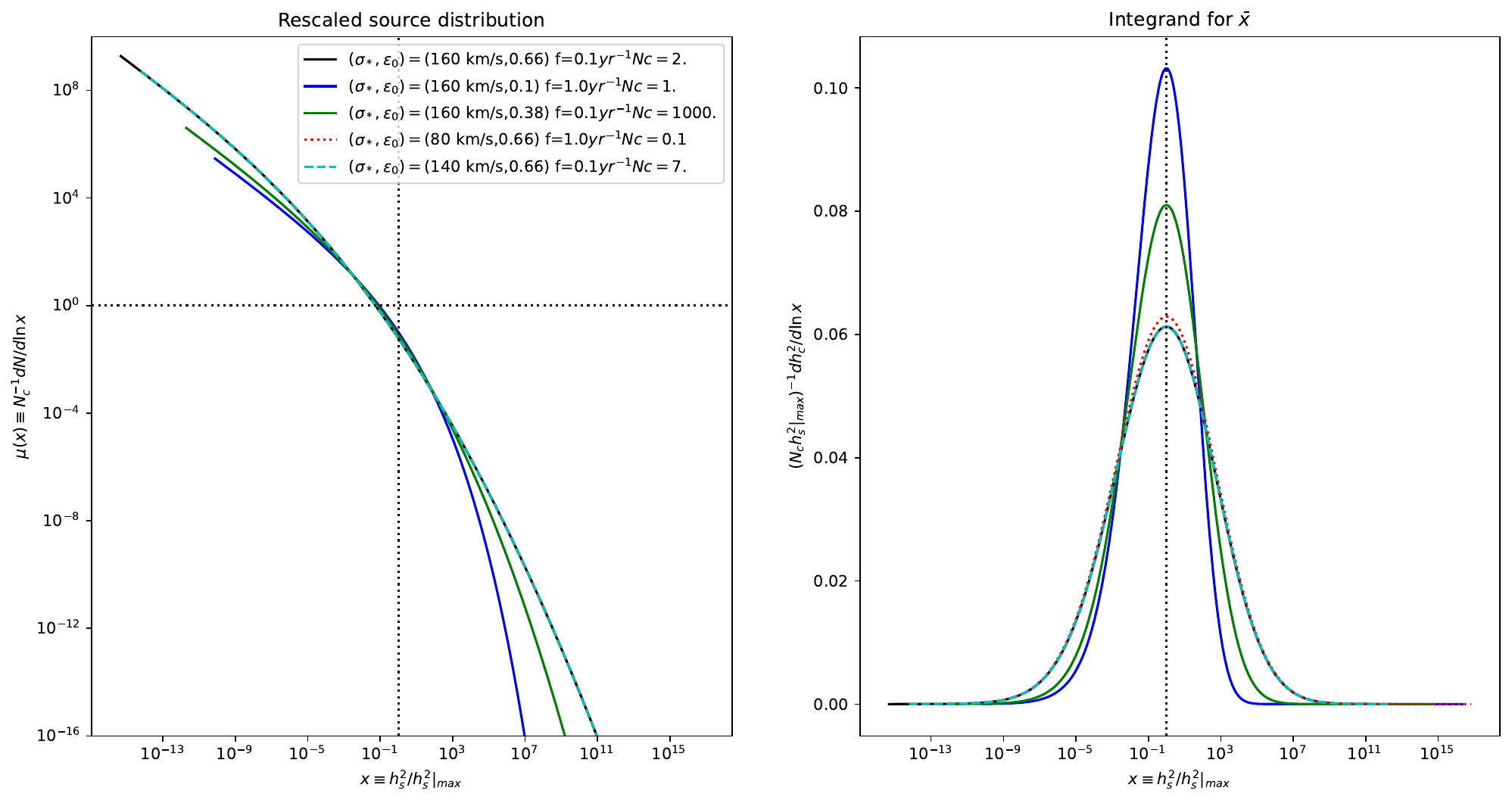}
    \caption{Examples of $\mu(x)$ for several choices of $M_{\rm peak}$ and frequency. We change the peak mass by either changing $\sigma_\star$ or the scatter in the $M-\sigma$ relation. One can see that the frequency dependence is completely reabsorbed by our scaling. When $M_{\rm peak}$ is changed by changing $\sigma_\star$ the shape of $\mu$ is also unchanged. Changing the scatter of the $M-\sigma$ relation leads to slightly different forms for $\mu$. We show with a black solid line, the fiducial $\mu$ we use in this appendix. In the main body of the paper we change the parameters of the $M-\sigma$ relation and compute the correct distribution for each case. The difference in shapes shown here are subdominant to the change in $h_s^2|_{\rm peak}$ and normalization of ${dN}/{d\log h_s^2}$ that the rescaling absorbs. In the cases shown both the normalization, $N_c$, and  $h_s^2$ change by approximately four orders of magnitude. The values for the five lines are:  $N_c=[2,1,10^3,10^{-1},7]$ and $h_s^2|_{\rm peak}=[10^{-28},2\times 10^{-31},3 \times 10^{-32}, 8 \times 10^{-32}, 9\times 10^{-30}]$. the corresponding values of $M_{\rm peak}$ are: $M_{\rm peak}=[4\times 10^{10},10^{9},3 \times 10^{9}, 8 \times 10^{9}, 2\times 10^{10}]$.} \label{fig:fid_mu}
\end{figure}

Figure \ref{fig:fid_mu} shows the fiducial $\mu(x)$ we will use together with the corresponding functions for a few representative model parameters to illustrate its sensitivity. These should be contrasted with the variation of $N_c$ which can change by many orders of magnitude as we change frequency or $M_{\rm peak}$ (equation \ref{eq:Nc_h2c_scaling}).

We consider the sum over sources of the scaled square of their strain, $x_t=\sum_s x_s$ and we will compute the corresponding $P(x_t)$ for different values of $N_c$. The mean satisfies:
\[
\bar x_t =\int x_t P(x_t) dx_t =\int x \frac{dN}{dx} dx = N_c.
\]
We will calculate $P(x_t)$ by  computing the Fourier transform of $P(x_t)$ and  by doing an FFT to go back to $x_t$ space. The values of $x$ in an FFT are equally spaced, $x_k=k x_{\rm min}$ for some $x_{\rm min}$ with $k$ an integer. One aspect that is very clear in figure \ref{fig:fid_mu} is that the range of $x$ that contribute to $x_t$ is extremely large and thus it is not possible to cover with a reasonably sized FFT. Doing such a brute force FFT would also be an overkill. 

One needs to choose the range of $x$ to cover with the numerical calculation. A first suggestion would be to try to cover the peak of the kernel for $x_t$ but this suggestion would not lead to an accurate calculation if $N_c$ differs from one by a lot. A driving consideration is that for numerous sources, the distribution of power is very well approximated by a Gaussian, so that part of the luminosity function need not be computed numerically. It is the contribution from sources that are not numerous that leads to a non-trivial shape for the distribution and needs to be computed numerically. 

In practice we will use as $x_{\rm min}$ for the FFT the value such that the characteristic number of sources below that point is a hundred. Any other large number would be equally good as long as it $x_{\rm min}$ is large enough that with the size of the FFT used the maximum value of $x$ is well in the tail, {\it ie.} $dN/dx \ll 1$. In the main sections of this work, we adopt $x_{\rm min} = h^2_p/h^2_s|_{\rm peak}$, where the pivot point was previously defined as the strain after which the number of sources is 1, which ensures an appropriate range of $x$ values for any model. Incorporating the contribution from the faint sources below $x_{\rm min}$ is easy, one needs to correct $u(\omega)$ by:
\[
\Delta u(\omega) = -1/2 \omega^2 \sigma^2 - i \Delta x \omega
\label{eq:gauss_approx}
\]
with 
\begin{align}
  \Delta x &=\int_0^{x_{\rm min}} dx\frac{dN}{dx} x   \nonumber \\
  \sigma^2 &=\int_0^{x_{\rm min}} dx\frac{dN}{dx} x^2.
  \label{eq:below_xmin}
\end{align}

As $N_c$ varies both $x_{\rm min}$ and the value of $x$ beyond which there is only one source will change.  The shape of $dN/dx$  in this range will evolve and as we learned in Appendix \ref{app:dist_power}, it is this shape what determines the shape of the distribution of $x_t$. 
 
\begin{figure}
    \centering
    \includegraphics[width=0.85\linewidth]{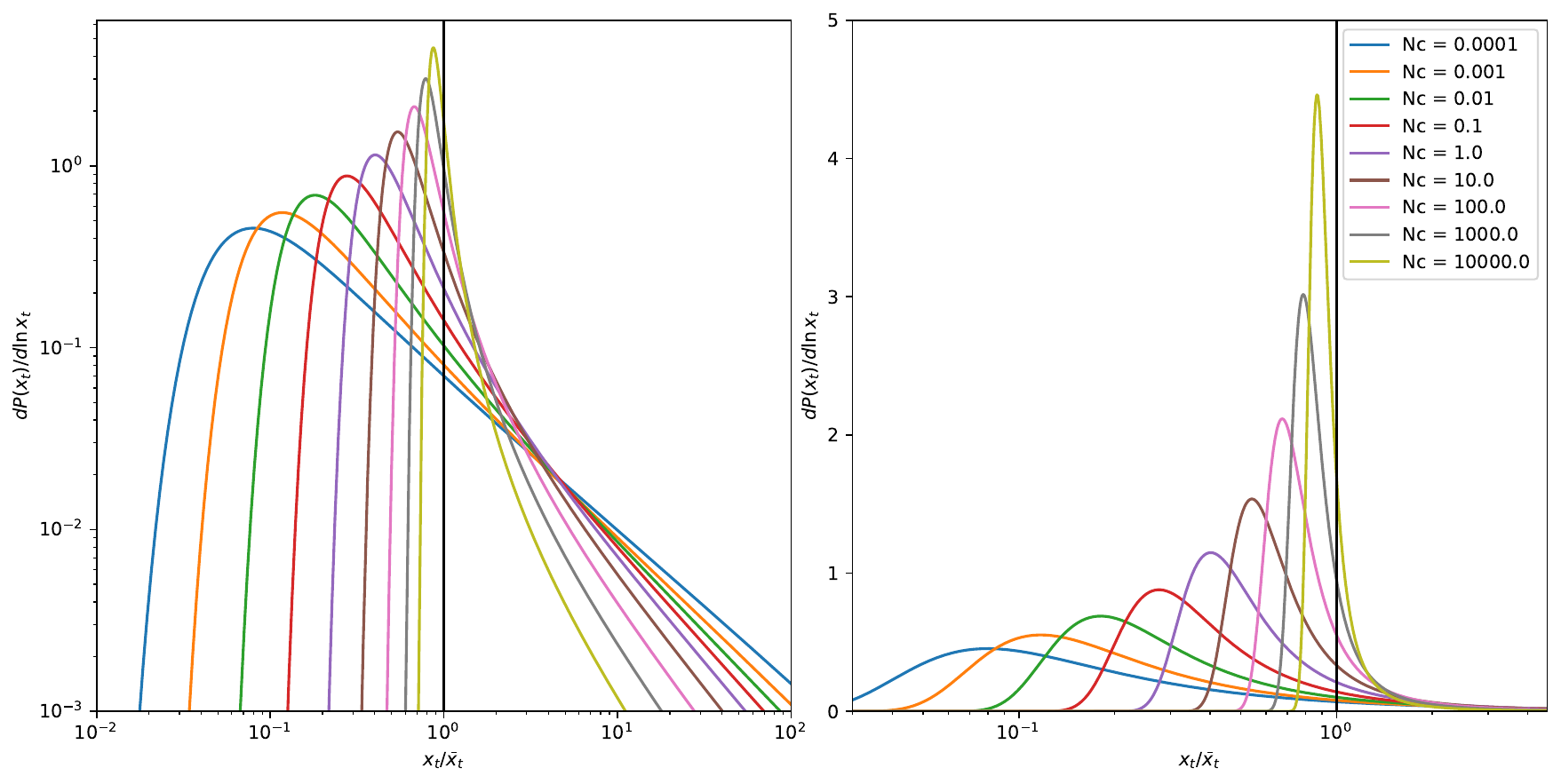}
    \caption{$P(x_t)$ for different values of $N_c$. We have normalized the $x$ axis by the mean $x_t$ of each case ($N_c$). We chose and FFT of size $2^{24}$ with $x_{\rm min}$ set to the value of $x$ below which there are one hundred sources on average. The contribution from sources below $x_{\rm min}$ was added using a Gaussian approximation (equation \ref{eq:gauss_approx}).  For low values of $N_c$ the peak of the distribution differs substantially from the mean (solid vertical line).  }
    \label{fig:distrib_vs_Nc}
\end{figure}

The resulting distribution $P(x_t)$ for varying values of $N_c$ are shown in figure \ref{fig:distrib_vs_Nc}. Several properties are apparent in the figure, although we have normalized the $x$ axis by the mean, the peak of the distribution shifts substantially. For small values of $N_c$ the mean is far in the tail of the distribution, while when $N_c$ is very large the mean and the peak coincide and the distribution is very close to a delta function.   

This behavior can be understood easily using the intuition we developed using the power law distributions. We know that the relevant value of $x$ is that which sets the transition between numerous and rare sources which we denote $x_1$. In the realistic distributions we are considering the slope of the luminosity function evolves crossing the special $\gamma=2$ value. When  $N_c \approx 1$, $x_1\approx 1$.  For very large $N_c\gg 1$ then  $x_1 \gg 1$ and in that range $\mu(x)$ is very red, it has $\gamma$ significantly larger than two. In that regime the sources much fainter than $x_1$ which are very numerous, dominate and the resulting distribution is very Gaussian and centered at $\bar x_t$.  In the opposite limit, when $N_c\ll 1$ one has $x_1\ll 1$ which falls in a part of the distribution where $\gamma$ is below two and thus the distribution becomes very non-Gaussian with the mean falling deeply in the tail. 

\begin{figure}
    \centering
    \includegraphics[width=0.5\linewidth]{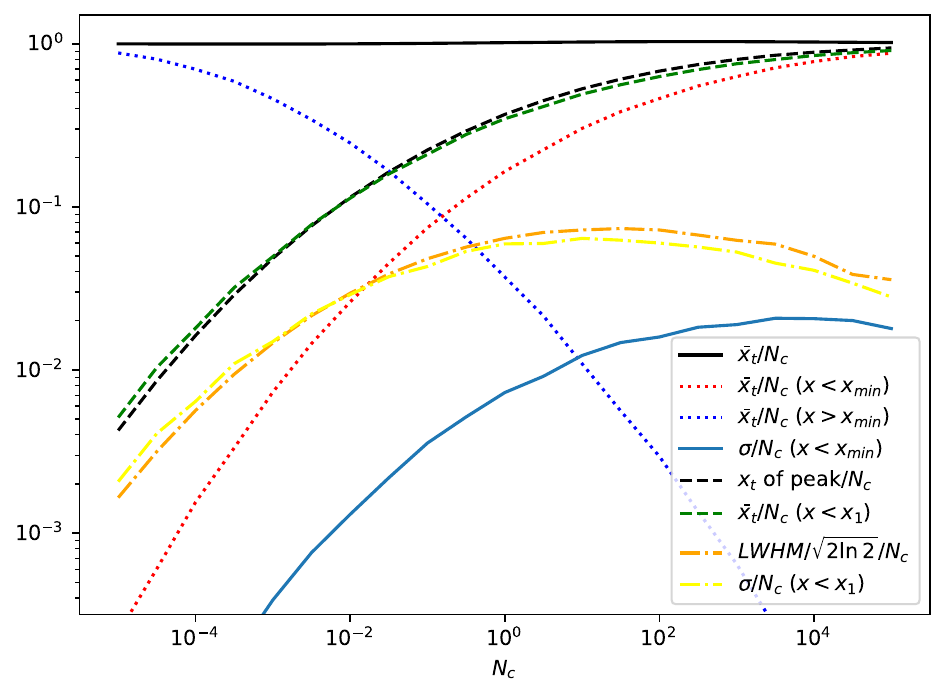}
    \caption{Summary of properties of $P(x_t)$ as $N_c$ changes. The mean of the distribution computed analytically is $N_c$. It was used to normalize all variables. We show the contributions to the mean from sources above and below $x_{\rm min}$, sampling used in the FFT. The contribution from sources below $x_{\rm min}$ is included using a Gaussian approximation. We show both the mean and standard deviation of this Gaussian. Although these sources provide the dominant contribution to the mean at high values of $N_c$, the associated variance is always small. We also show the evolution of the position of the peak of $P(x_t)$ which agrees very well with the estimate in equation \ref{eq:peak-approx}. The width of the Gaussian-like core is quantified using width at half maximum to the left of the peak ($LWHM$)  which is shown to agree well with the estimate in equation \ref{eq:sigma-approx}.}
    \label{fig:summary_dist_vs_Nc}
\end{figure}

In figure \ref{fig:summary_dist_vs_Nc} we show several properties of $P(x_t)$ as $N_c$ changes. The mean of the distribution is $\bar x_t=N_c$ so we will scale all results by $N_c$. Both sources below and above $x_{\rm min}$ contribute to this mean and their relative contribution evolves with $N_c$. At low values values of $N_c$ the bright sources dominate while the opposite is true once $N_c$ gets large.  

Figure \ref{fig:summary_dist_vs_Nc} also shows the properties of the Gaussian-like core of the distribution. The peak evolves with $N_c$ and starts well below the mean for $N_c\ll 1$, but as $N_c$ grows the peak and the mean of the distribution become ever closer. We can also see in \ref{fig:summary_dist_vs_Nc} that the peak of the distribution is approximately given by the mean predicted by the numerous sources: 
\[
x_{\rm peak}\approx \int_0^{x_1} dx \frac{dN}{dx} x.
\label{eq:peak-approx}
\]

To characterize the with of the Gaussian-like core we can consider the part of the distribution to the left of the peak, as the part to the right is affected by the presence of the tail. We can quantify the width by measuring the distance between the peak and the place where the $P$ is half the peak value, the which we can call left width at half max ($LFHM$).  If the distribution was a Gaussian the $LWHM$ would be given in terms of the standard deviation $\sigma$ as $LWHM = \sigma_{\rm peak} \sqrt{2 \log 2}$. As we did for the position of the peak, we can estimate $\sigma$ using the contribution from the numerous sources only:
\[
\sigma^2_{\rm peak}\approx \int_0^{x_1} dx \frac{dN}{dx} x^2.
\label{eq:sigma-approx}
\]
Figure \ref{fig:summary_dist_vs_Nc} shows that this is an excellent estimate. 

Finally, we recall that the contribution from the sources below $x_{\rm min}$ is modelled as a Gaussian with mean and standard deviation given by equation \ref{eq:below_xmin}.  Figure \ref{fig:summary_dist_vs_Nc} shows both mean and standard deviation. One can see that except at the very largest value of $N_c$ for our choice of $x_{\rm min}$ the standard deviation of this Gaussian is much smaller than the width of $P$. Thus the sources below $x_{\rm min}$ can shift the mean but have little effect on the width. This would be even more true if were to pick a smaller $x_{\rm min}$. 

\section{Multiple mergers}\label{app:multiple-gen-mergers}

In this appendix we consider the contributions of multiple generation of mergers. To gain some intuition  we assume for simplicity all mergers are equal mass. This simple case will illustrate what the expected behavior of the power distribution is and highlight the fact that it will mainly result in a modest shift towards higher backgrounds but will leave the properties of the tail of the distribution unchanged. 

If all mergers are equal mass the quantities describing the strain and the characteristic number of sources will evolve as:
\begin{align}
  h_{c,N}^2 &= 2^{-2(N-1)/3} h_{c,1}^2 \nonumber \\
  h_{s,N}^2 &= 2^{-10(N-1)/3} h_{s,1}^2 \nonumber \\
  N_{c,N} &= 2^{8(N-1)/3} N_{c,1},
\end{align}
where $N$ is denotes the generation with $N=1$ being the last merger.  We can see that $N_c$ increases rapidly making the distribution of the total strain contributed by each generation become narrower. At the same time  the average strain decreases. 

\begin{figure}
    \centering
    \includegraphics[width=0.85\linewidth]{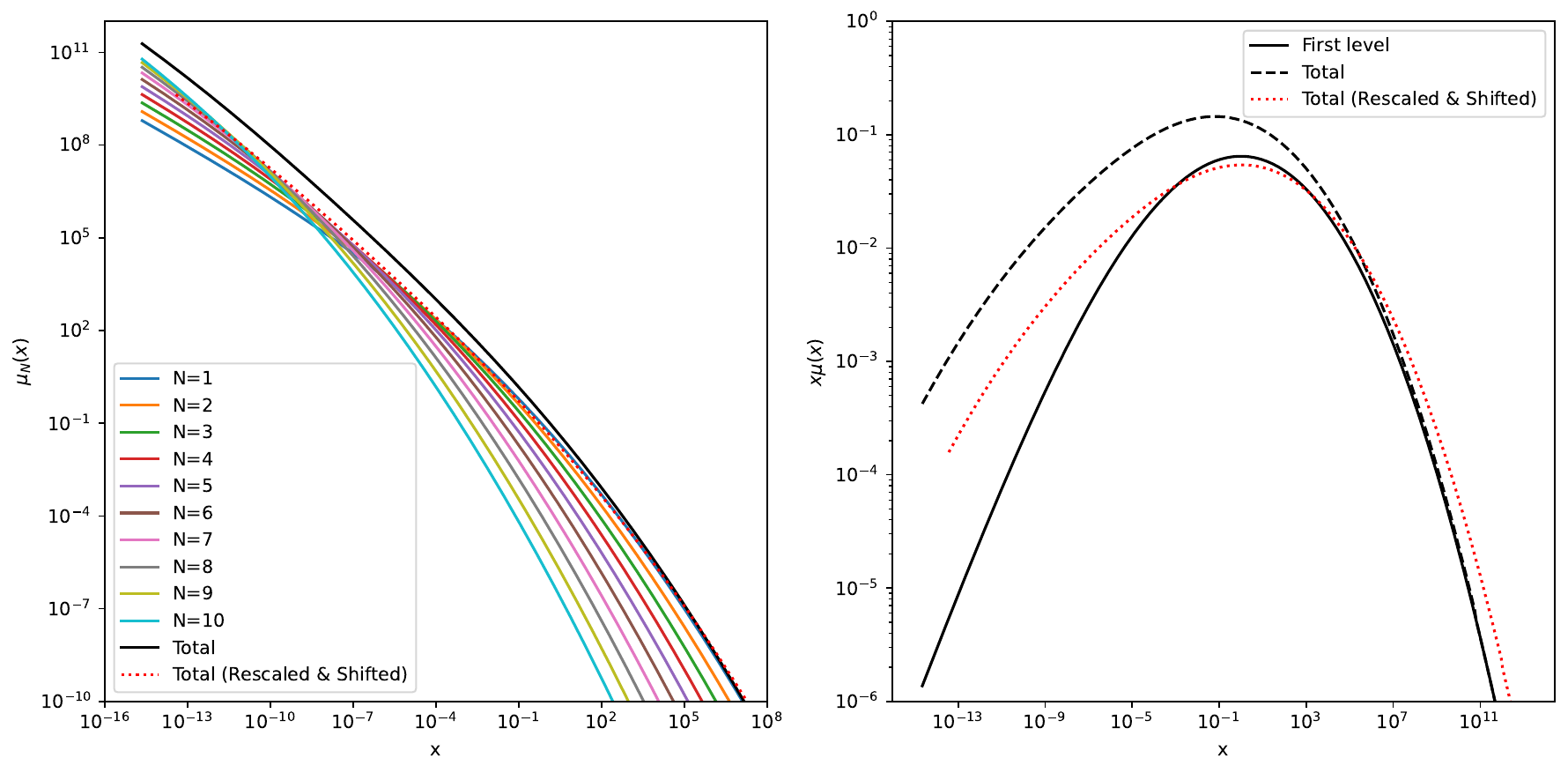}
    \caption{Contributions from each generation of mergers as well as the total contribution $\mu_T=\sum \mu_N$. With this definition $x=1$ does not correspond to the peak of $\mu_T$ and $\mu_T$ is not normalized to one. We also show the rescaled version of $\mu_T$. We note that for $x\gg 1$, $\mu_T\approx \mu$ because one gets bright sources almost exclusively from the last generation of mergers.  For $x\ll 1$, however, previous mergers make a significant contribution and make the slope of $\mu_T$ steeper. For high values of $N_c$ where this part of the distribution ($x\ll 1$) controls the shape of the $P(x_t)$ this steeper slope with result in a distribution where the peak is closer to the mean. }
    \label{fig:mu-multiple-gens}
\end{figure}

One can compute $\mu_N(x)$ defined as the contribution from each of the levels of the tree,
\[
\frac{dN}{d\log x} = N_{c,1} \mu_T(x) \ \ , \ \ x=\frac{h_s^2}{h_s^2|_{peak,1}},
\]with $\mu_T=\sum \mu_N$. Notice that we have used the values for the last set of mergers to normalize our variables. With these definitions:
\[
\mu_N(x)=2^{8(N-1)/3} \mu_1(2^{10(N-1)/3}\ x).
\]
Figure \ref{fig:mu-multiple-gens} shows the different contributions $\mu_N$ as well as the total $\mu_T$. As a result of the sum, the overall amplitude increases and the peak of $x\mu(x)$ is shifted to the left. 

It is important to emphasize that because we have used the peak strain of the last generation of mergers, with this definition $\mu_T(x)$ does not integrate to one and its peak is not at $x=1$. It is perhaps more convenient to shift and renormalize  $\mu_T$. The rescaled version of $\mu_T$ is also shown in figure   \ref{fig:mu-multiple-gens}. If one thinks in this way, adding all the levels corresponds to a rescaling of the amplitude of the mass function and a shift of the peak mass. The right panel of $\ref{fig:mu-multiple-gens}$ shows that the peak for $x \mu_T(x)$ happens for an $x$ smaller than one $(\approx 0.06)$, and thus the distribution has a lower $M_{\rm peak}$ $(\propto x^{3/10})$. One can also see from the left panel that the normalization is higher (the integral is 2.67 rather than one) which implies a higher effective $\phi_\star$. For this simple case the scalings are:
\begin{align}
    \phi_\star \rightarrow & 2.67 \times \phi_{\star} \nonumber \\
    M_{\rm peak} \rightarrow&  0.43 \times M_{\rm peak} 
\end{align}

\begin{figure}
    \centering
    \includegraphics[width=0.85\linewidth]{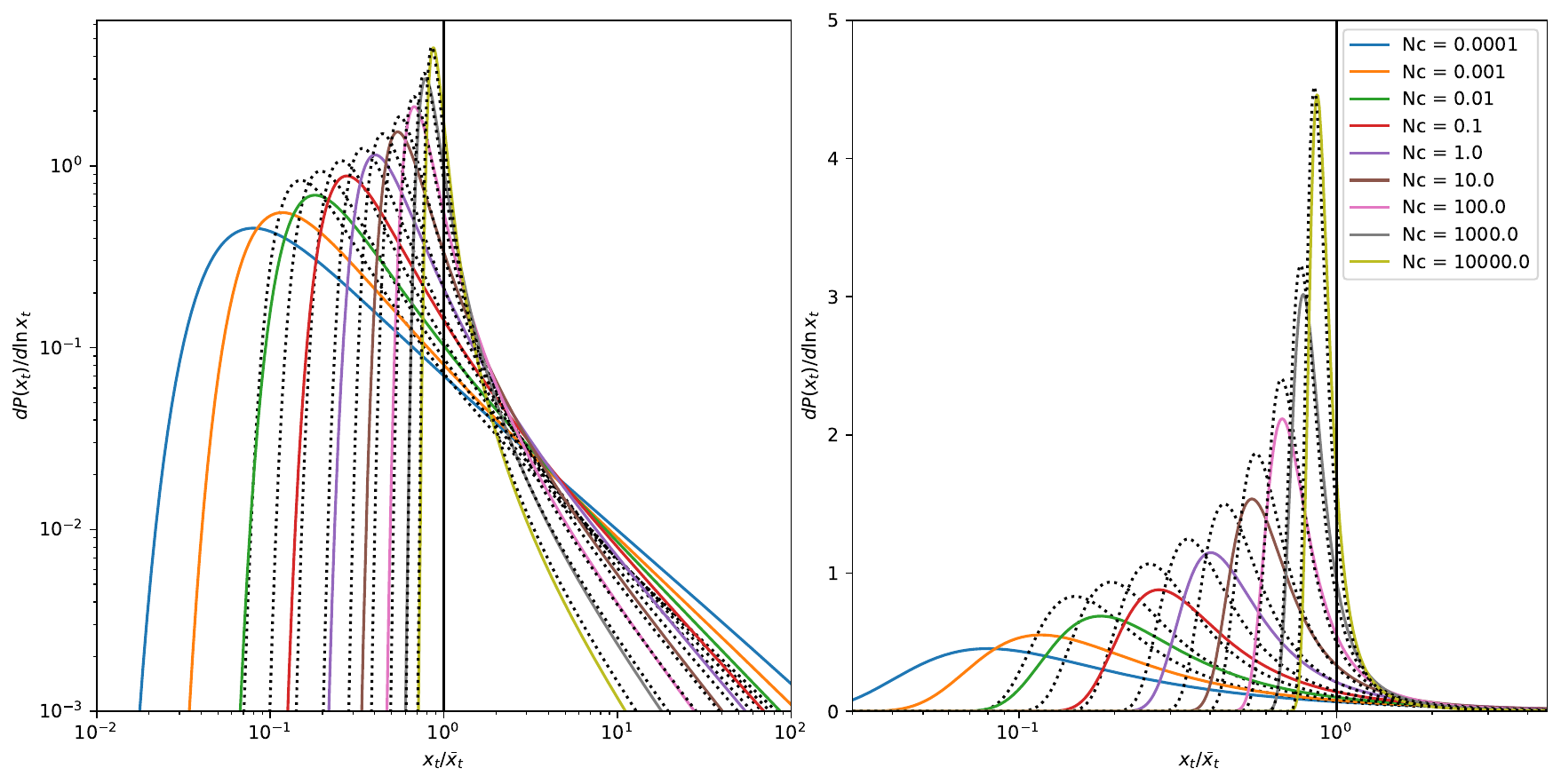}
    \caption{$P(x_t)$ for different values of $N_c$. The dotted curves show the corresponding values when multiple generations of sources are included.  The solid curves are the same as in figure \ref{fig:distrib_vs_Nc}.  For low values of $N_c$,  $x_1 \ll 1$, where $\mu_T$ has a steeper slope compared to $\mu$. As a result the difference between the peak of $P(x_t)$ and its mean will be smaller that when one generation is considered. The   of $P(x_t)$ will also be steeper.  For large values of $N_c$, $x_1\gg 1$,  a range of  $x$ where higher generations of mergers do not contribute much to $\mu_T$ and thus $P(x_t)$ is barely affected. }
    \label{fig:P-multiple-levels}
\end{figure}

In addition to the shift of the peak, one can see that the slope of $\mu_T(x)$ below the peak is steeper than for the case with only one generation of mergers. This will increase the relative contribution of the numerous faint sources when we compute the distribution of the overall strain. A comparison between the single level prediction and the multiple level one is shown in figure \ref{fig:P-multiple-levels}. The difference is all coming from the faint sources below the peak contribution to $x\mu(x)$. As a result for low values of $N_c$ where $x_1 \ll 1$, the steeper slope implies that the difference between the peak of $P(x_t)$ and the mean will be smaller. It will also result in a steeper tail for $P(x_t)$.  For large values of $N_c$, $x_1\gg 1$, where higher generations of mergers do not contribute much to $\mu$ and thus $P(x_t)$ is barely affected. 

\end{document}